\documentclass{article}
\PassOptionsToPackage{numbers,square,comma,sort&compress}{natbib}
\usepackage[main,final]{neurips_2026}
\usepackage[utf8]{inputenc}
\usepackage[T1]{fontenc}
\usepackage{amsmath,amssymb,booktabs,graphicx,microtype,array,url,wrapfig,enumitem}
\makeatletter\newcommand\captionof[1]{\def\@captype{#1}\caption}\makeatother
\usepackage[table]{xcolor}
\usepackage{flafter}
\usepackage{tikz}
\usepackage{hyperref}
\makeatletter\g@addto@macro\UrlBreaks{\do\-}\makeatother
\hypersetup{colorlinks=true,linkcolor=blue!45!black,citecolor=blue!45!black,urlcolor=blue!45!black}
\setcitestyle{numbers,square,comma,sort&compress}
\makeatletter\newcommand{\FloatBarrier}{\par\ifx\@deferlist\@empty\else\newpage\fi}\makeatother
\newcommand{\tableinput}[1]{\csname @@input\endcsname #1 }
\newcolumntype{P}[1]{>{\raggedright\arraybackslash}p{#1}}
\newcommand{\yes}{$\checkmark$}
\newcommand{\no}{$\times$}
\definecolor{c2shade}{HTML}{E3EEF8}
\definecolor{gateshade}{HTML}{EAF4FB}
\definecolor{verdictshade}{HTML}{EAF7EE}
\definecolor{promptbg}{RGB}{246,246,250}
\newcommand{\codeurl}{\url{https://github.com/koanatakiyo/SocCogEval}}

\title{Social Interaction Breaks Replicate Independence in Controlled-Clone LLM Agents}
\author{%
  \textbf{Yandan Zheng}$^{1,3}$, \textbf{Haoran Luo}$^{1}$\thanks{Corresponding author.}, \textbf{Anh Tuan Luu}$^{1,2}$\\
  $^{1}$Nanyang Technological University, Singapore \quad $^{2}$VinUniversity, Vietnam\\
  $^{3}$Alibaba-NTU Global e-Sustainability CorpLab (ANGEL), Singapore\\
  \texttt{\{yandan.zheng,haoran.luo,anhtuan.luu\}@ntu.edu.sg}
}

\begin{document}
\maketitle
\begin{abstract}
Multi-agent LLM evaluations and social simulations create many agents from one model and treat them as independent replicates, even after the agents interact. This assumption remains untested with matched controls. We introduce SocCogEval, a controlled-clone assay for Social Cognitive Transfer (SCT). Clones share the base model, system prompt, and decoding parameters and differ only in a private brief, and a factorial protocol separates the brief, 30 rounds of interaction, and initialization noise. On Qwen3-122B-FP8, interaction with the briefs held fixed raises off-topic BFI-10 response-profile drift by $\bar{\Delta}{=}{+}0.917$ and ${+}0.912$ in two topically disjoint scenarios (exact $p{=}0.0008$ and $0.0039$). The drift is larger than under the brief alone, persists under temperature-zero re-probing, and stays positive without typed memory extraction in exploratory tests. Matched solo generation reproduces part of it, and social interaction remains highest. Across five more model variants, every scenario-averaged interaction contrast is positive. Each interacting population is therefore one replicate, and treating its agents as independent requires paired no-interaction controls. Our code and data are publicly available.\footnote{\raggedright \codeurl}
\end{abstract}

\section{Introduction}\label{sec:intro}

\begin{wrapfigure}[14]{r}{0.46\textwidth}
\vspace{-12pt}
\centering
\includegraphics[width=\linewidth]{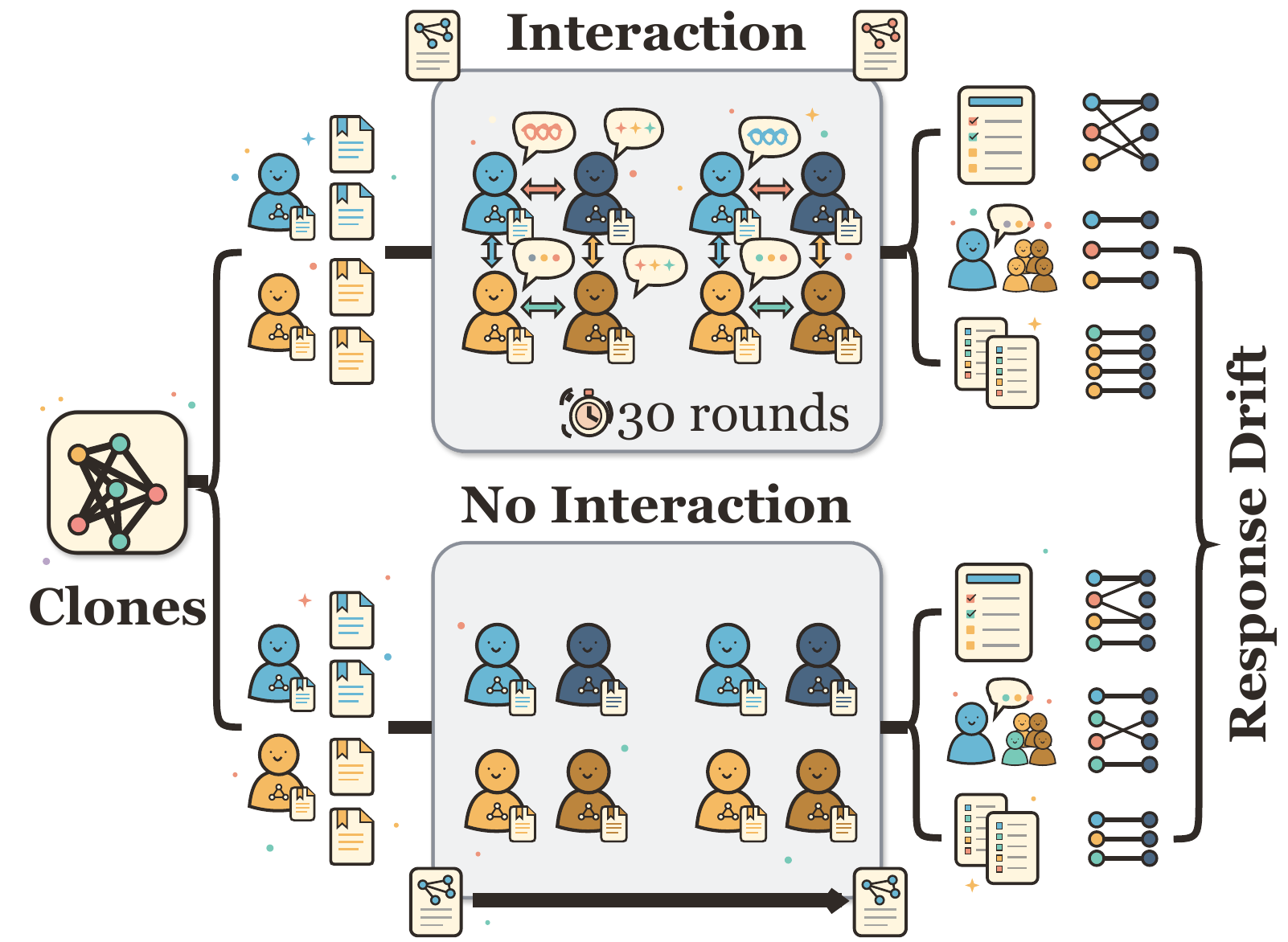}
\caption{\textbf{Interaction moves clones:} same briefs, with or without 30 rounds.}\label{F0}
\vspace{-10pt}
\end{wrapfigure}

Many LLM systems build a population from copies of one model, including multi-agent debate~\citep{Du2024_MultiagentDebate,Liang2024_DivergentDebate,Du2024_Debate,Smit2024_MAD}, collaboration and evaluation frameworks~\citep{Li2023_CAMEL,Wu2024_AutoGen,Hong2024_MetaGPT,Chan2024_ChatEval}, simulated survey samples~\citep{Argyle2023,Aher2023_SimulateHumans}, and simulations of conflict, opinion, cooperation, and whole societies~\citep{Hua2024_WarAgent,OASIS,AgentSociety,Generative_Agents,Park2024_1000,Concordia,Li2024_SocioVerse,Chuang2024_OpinionDynamics,CisnerosVelarde2025_Funding,Piatti2024_GovSim,FlintAshery2025_Conventions,Anthis2025_SocialSim}. These studies treat the copies as comparable replicates of one system, even after many rounds of talk. If interaction changes the agents, each interacting population is one sample, and counting agents or posts as independent observations overstates the evidence~\citep{Hurlbert1984,Miller2024_ErrorBars}.

\begin{figure}[t]
\centering
\includegraphics[width=\textwidth]{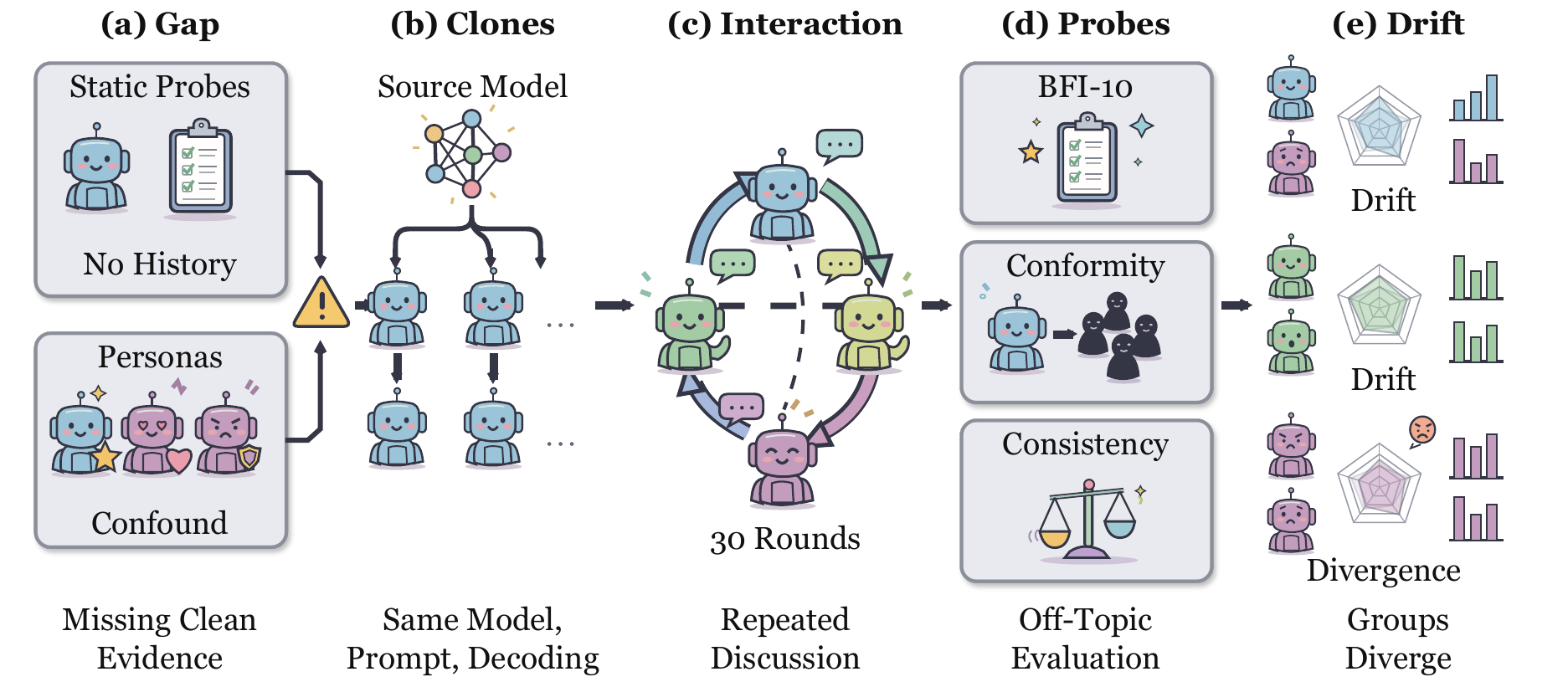}
\caption{\textbf{SocCogEval at a glance:} the gap, controlled clones, interaction, off-topic probes, drift.}\label{F1}
\end{figure}

This assumption still lacks a controlled test (Figure~\ref{F1}). Benchmarks score a frozen model on curated items or a short episode~\citep{MT-bench,AgentBench,ToMBench,Kim2023_FANTOM,Xu2024_OpenToM,CogBench,CogMir,PsychoBench}. Social simulations run long interactions but assign personas or opinions at the start, which mixes injected priors with the effect of interaction. Studies of drift, stability, and conformity measure the discussed task or the dialog itself~\citep{Li2024_PersonaDrift,Laban2025_LostMultiTurn,Kovac2024_StickToRole,Weng2025_Conformity,Potter2026,Rath2026_AgentDrift}, and personalities can emerge when undifferentiated agents communicate~\citep{Takata2024}. A controlled test has to meet three challenges. It must inject no differences at initialization, separate the effect of a brief from the effect of interaction, and use a probe unrelated to the discussed topic, analyzed at the level of the run.

SocCogEval meets them with controlled clones that share base model, system prompt, and decoding parameters, each holding at most one private brief. The Factorial Social Interaction Protocol (FSIP) crosses brief presence with 30 rounds of interaction, so the pure-interaction contrast $(C_2){-}(C_4)$ holds the brief fixed and toggles interaction (Figure~\ref{F0}). A solo-generation condition $C_6$ matches the call budget without any partner. The measurements follow three nested questions: do clones diverge on the discussed decision, do they predict exposed in-group peers better, and does the divergence transfer to an off-topic BFI-10 response profile?

\begin{figure}[t]
\centering
\includegraphics[width=\textwidth]{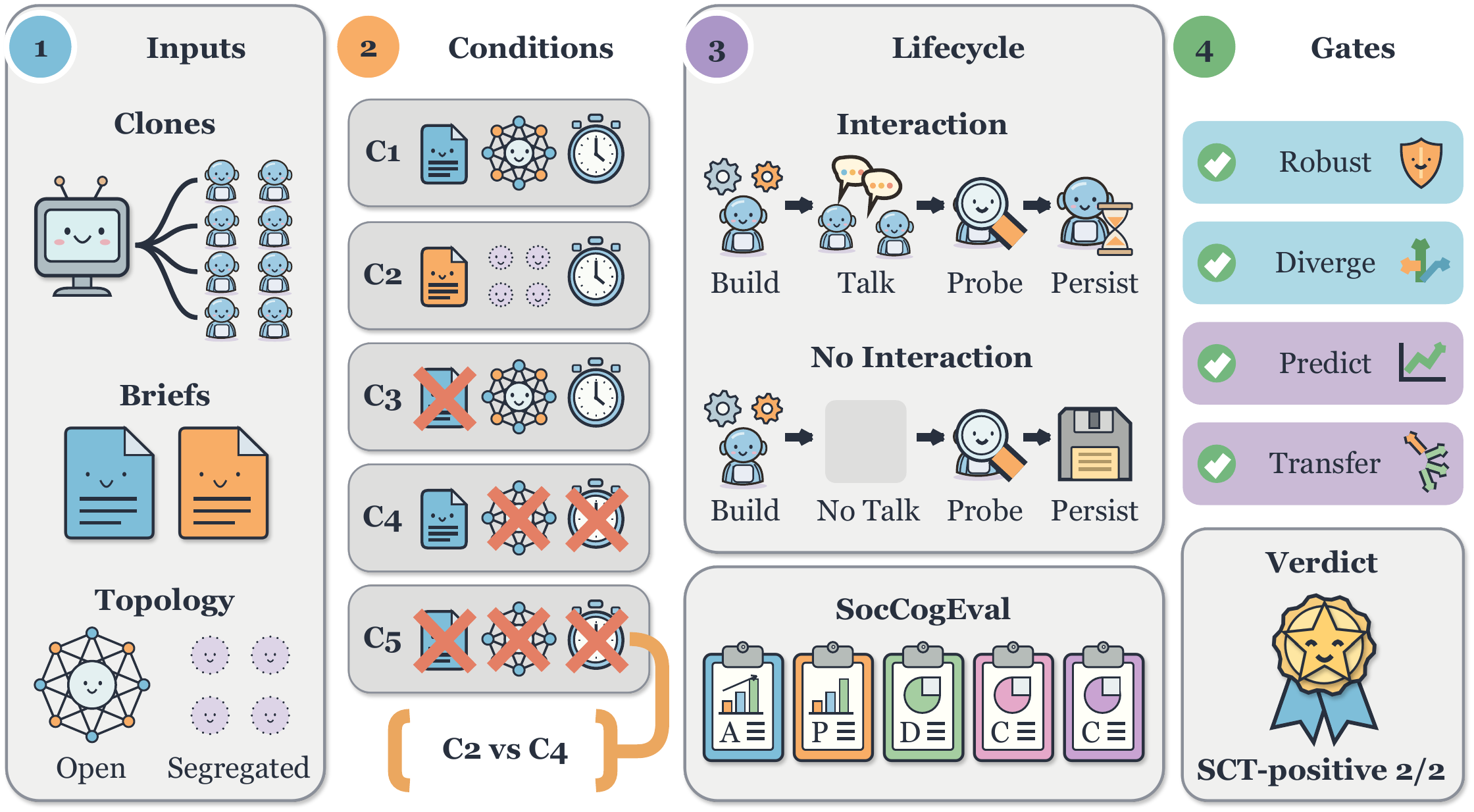}
\caption{\textbf{FSIP pipeline:} inputs, conditions $C_1$--$C_5$, run lifecycle, gates and verdict.}\label{F2}
\end{figure}

Social interaction breaks replicate independence on the primary transfer outcome. On Qwen3-122B-FP8, the pure-interaction contrast is $\bar{\Delta}{=}{+}0.917$ BFI-10 L1 units on Greenfield and ${+}0.912$ on Riverton (run-level $p{=}0.0008$ and $0.0039$, $g{\approx}1.0$), about 0.18 Likert points per factor. It tracks interaction rather than the brief, and it survives deterministic re-probing and a verbatim-history ablation that removes typed memory extraction. Solo generation reproduces part of the shift, while social interaction remains highest in both scenarios. Five more model variants give positive but uneven breadth, and all five have positive scenario-averaged $(C_2){-}(C_4)$ contrasts.

Our contributions are fourfold. (i) We define the Social Cognitive Transfer (SCT) task with three nested sub-tasks, Decision, Prediction, and Transfer. (ii) We propose FSIP, a five-condition design that isolates initialization noise, brief alone, interaction given brief, and brief given interaction, with a matched solo-generation condition $C_6$ as the mechanism boundary. (iii) We introduce SocCogEval, which combines task probes, a primary BFI-10 transfer instrument~\citep{Rammstedt2007_BFI10,PsychoBench}, secondary conformity and consistency probes~\citep{Asch1956_Conformity,CogMir}, and a contrast schema that turns measurements into causal claims. (iv) We show that social interaction breaks replicate independence on the primary transfer outcome, with cross-scenario stability and positive cross-model direction. The reported drift is restricted to Part 2 instruments unrelated to the discussion and must clear a three-part criterion (Section~\ref{sec:metrics}). SocCogEval is an identification protocol for post-interaction response-profile drift (Section~\ref{sec:exp-limits}).

\section{Related work}\label{sec:related}

\textbf{Social-cognition and psychometric evaluation.} ToMBench~\citep{ToMBench}, FANToM~\citep{Kim2023_FANTOM}, OpenToM~\citep{Xu2024_OpenToM}, CogMir~\citep{CogMir}, CogBench~\citep{CogBench}, and PsychoBench~\citep{PsychoBench} score frozen models on curated items. Psychometric frameworks measure and shape LLM personality~\citep{SerapioGarcia2025_PsychometricLLM,Ye2025_LLM_Psychometrics}, expressed values shift with context~\citep{Kovac2024_StickToRole}, and the reliability of self-report inventories for LLMs is contested~\citep{Binz2023,Shu2024_PsychReliability,Gupta2024_SelfAssessment,Huang2024_ScaleReliability,Lee2025_TRAIT,Jiang2024_PersonaLLM,Han2025_PersonalityIllusion}. These works measure a snapshot and cannot separate what is in the weights from what an interaction history adds. SocCogEval places BFI-10~\citep{Rammstedt2007_BFI10,PsychoBench} and Asch-style items~\citep{Asch1956_Conformity} in a paired pre-post design, so scores become difference-in-differences statistics that read as transfer.

\textbf{Agent societies and multi-agent systems.} OASIS~\citep{OASIS}, AgentSociety~\citep{AgentSociety}, generative-agent worlds~\citep{Generative_Agents,Park2024_1000}, Concordia~\citep{Concordia}, SocioVerse~\citep{Li2024_SocioVerse}, and WarAgent~\citep{Hua2024_WarAgent} simulate LLM populations in which opinions~\citep{Chuang2024_OpinionDynamics,CisnerosVelarde2025_Funding}, cooperation~\citep{Piatti2024_GovSim}, conventions~\citep{FlintAshery2025_Conventions}, and individual personalities~\citep{Takata2024} emerge. SOTOPIA scores social skill within an episode~\citep{Zhou2024_SOTOPIA}, and LLM simulation is proposed as a social-science method~\citep{Anthis2025_SocialSim,Mou2024_SocialSimSurvey}. Debate, evaluation, and collaboration frameworks build teams from one model~\citep{Du2024_MultiagentDebate,Liang2024_DivergentDebate,Du2024_Debate,Smit2024_MAD,Chan2024_ChatEval,Li2023_CAMEL,Wu2024_AutoGen,Hong2024_MetaGPT}, often with persistent memory~\citep{MemGPT,Xu2025_AMEM,Mem0}. These studies inject heterogeneous priors or report collective outcomes, so they cannot isolate what interaction does to each agent. Controlled clones and a factorial contrast schema remove that confound.

\textbf{Drift, stability, and conformity.} Instructions fade within a dialog~\citep{Li2024_PersonaDrift}, reliability drops over multi-turn conversations~\citep{Laban2025_LostMultiTurn}, models agree with interlocutors~\citep{Sharma2024_Sycophancy} and conform to majorities~\citep{Weng2025_Conformity}, multi-agent systems fail through coordination errors~\citep{Cemri2025_MASFailures}, and a peer's presence changes behavior~\citep{Potter2026}. Rath~\citep{Rath2026_AgentDrift} defines drift as progressive degradation of task quality, coordination, and inter-agent coherence through a twelve-component Agent Stability Index. These works measure instability on the task or in the dialog. We instead test distributional exchangeability of bit-identical clones. Initialization and brief assignment stay fixed, and we ask whether interaction transfers to off-topic response profiles under matched brief-only and no-interaction controls.

\textbf{Uncertainty in evaluation.} Observations within one unit are not independent replicates~\citep{Hurlbert1984}, shared-source items need clustered errors~\citep{Miller2024_ErrorBars,Cameron2015}, and synthetic samples can understate variation~\citep{Bisbee2024}. Multi-turn agent evaluation mostly targets task success~\citep{Yao2023_ReAct,MT-bench,AgentBench}. The run is our unit of replication.

\begin{figure}[t]
\centering
\includegraphics[width=\textwidth]{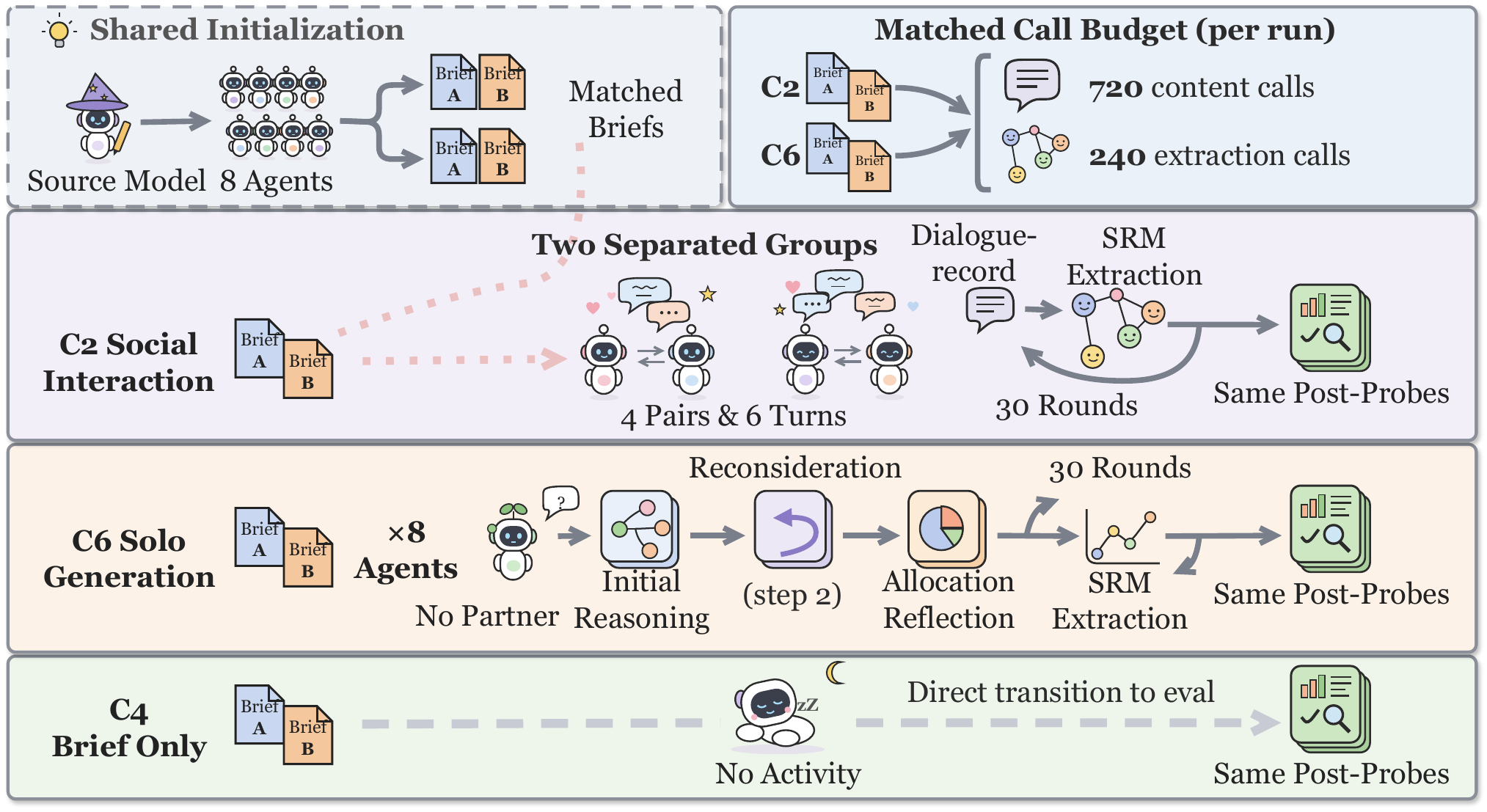}
\caption{\textbf{Three matched lanes:} shared start, social $C_2$, solo $C_6$ (same budget), brief-only $C_4$.}\label{F3sch}
\end{figure}

\section{SocCogEval: task, protocol, and measurement}\label{sec:method}\label{sec:task}

SocCogEval combines a task definition, a factorial protocol, and a measurement suite. The task defines what counts as social cognitive transfer. The protocol separates interaction from briefs, initialization noise, and memory. The measurements convert each run into contrast-level evidence.

\subsection{Task: Decision, Prediction, Transfer}\label{sec:sct-task}

An SCT instance specifies a frozen base model $M$, an agent population $A{=}\{a_1, \dots, a_k\}$ of $k$ clone agents that share system prompt and decoding parameters, an interaction graph $G$ over $A$ (all-to-all, two-component segregated, or empty), a scenario $S$ with a 5-dim allocation space, an optional brief assignment $T$ (asymmetric A and B, symmetric A, or none), and an interaction horizon $N$ (default 30 rounds). The design rests on the clone constraint. In any condition without briefs, $a_i$ and $a_j$ have bit-identical initial state. When briefs are assigned, the only initial difference is the private brief document. Three nested questions follow. SCT-Decision asks whether post-interaction allocation in segregated, asymmetrically briefed conditions diverges beyond what brief alone or initialization noise produce. SCT-Prediction asks whether agents in segregated conditions predict in-group members more accurately than out-group members. SCT-Transfer asks whether the divergence appears on instruments topically unrelated to the discussion, after subtracting the no-interaction baseline.

\subsection{Factorial Social Interaction Protocol}\label{sec:fsip}

The factorial axes are brief ($T \in \{\text{none}, \text{A only}, \text{asymmetric A and B}\}$) and interaction ($N \in \{0, 30\}$ with topology $G \in \{\text{open}, \text{segregated}\}$). The design uses five conditions plus one mechanism control (Figure~\ref{F2}). $C_1$ (open, symmetric brief) uses an all-to-all topology in which every agent receives Brief A, as an open-network reference. $C_2$ (segregated, asymmetric brief) splits agents into two components, half with Brief A and half with Brief B, and is the headline condition. $C_3$ (segregated, no brief) keeps the $C_2$ topology without briefs and isolates interaction alone. $C_4$ (brief only) assigns the A and B briefs with $N{=}0$ and isolates brief alone. $C_5$ (no brief, no interaction) sets the initialization noise floor. $C_6$ (solo generation) keeps the $C_2$ briefs, memory policy, probe cadence, and logical call budget, with private generation in place of partners (Figure~\ref{F3sch}).

The main causal contrast is $(C_2){-}(C_4)$ on every measured outcome. It holds the brief fixed, since both conditions use asymmetric A and B briefs, and toggles 30 rounds of interaction on or off. Interaction-created history, memory, and elapsed conversation are post-treatment pathways through which the treatment acts, so $(C_2){-}(C_4)$ is the total effect of interaction given the same brief in this protocol. $C_1$ uses symmetric briefs while $C_2$ uses asymmetric briefs, so the $(C_2){-}(C_1)$ contrast varies brief-symmetry and topology jointly and does not isolate topology. We therefore report $(C_2){-}(C_1)$ only descriptively and make no topology claim. Replica counts per condition are $\{C_1{=}7, C_2{=}8, C_3{=}5, C_4{=}5, C_5{=}3, C_6{=}5\}$ per scenario, with $k{=}8$ agents per replica.

\enlargethispage{\baselineskip}\textbf{Solo-generation control.} $C_6$ gives each agent three private generation calls per round (initial reasoning, reconsideration, allocation reflection) and one memory extraction, with no partner identity, messages, or feedback. This yields 720 content and 240 extraction calls per run, matching $C_2$'s intended logical totals. Its design was frozen before formal outcomes were examined, and all ten formal $C_6$ runs passed audit (Appendix~\ref{app:c6}). $(C_6){-}(C_4)$ measures the solo-generation shift, and $(C_2){-}(C_6)$ compares social and solo histories.

\subsection{Scenarios, runtime cycle, and models}\label{sec:fsip-cycle}

We run two topically disjoint scenarios. Greenfield is a community-grant allocation problem with \$10M across five categories such as Education, Healthcare, and Infrastructure. Riverton is a workforce transition budget problem with \$5M across categories such as Automation and Workforce Training. Both use 5-dim allocation vectors so Part 1 metrics are commensurable. Both have softened A and B briefs that frame the same problem through moderately divergent value lenses (Appendix~\ref{app:scenarios}). Per-round mechanics follow a Build-Talk-Probe-Persist cycle. Build updates a typed Social-Relational Memory (SRM) hypergraph from the previous round's conversations. Talk forms 4 random pair conversations among 8 agents with 6 turns each. Probe runs SocCogEval instruments at predetermined cadences, with belief and common-ground probes every five rounds that leave conversation history and memory unchanged. Persist writes the round artifacts to disk.

\textbf{Models and filtering.} The primary model is Qwen3-122B-FP8~\citep{Qwen2025_Qwen3} served locally via vLLM. Five model variants run on $\{C_2, C_4, C_5\}$ in both scenarios, three runs per condition and scenario: Qwen3-VL-32B-Instruct~\citep{Qwen2025_Qwen3VL} run text-only, DeepSeek-V3~\citep{DeepSeekAI2024_DeepSeekV3}\footnote{Historical DeepSeek $C_2$ runs used the provider identifier \texttt{deepseek-chat} and its $C_4$ runs used \texttt{deepseek-v4-flash}. Exact provider revisions were not recorded (Appendix~\ref{app:routing}).}, GPT-4.1-mini~\citep{OpenAI2025_GPT41}, Gemini-2.5-Flash~\citep{GoogleDeepMind2025_Gemini25}, and Llama-4-Scout~\citep{MetaAI2025_Llama4}. Figures abbreviate the six models as Q-122B, Q-VL, DS-V3, GPT-mini, Gem-Flash, and L4-Scout. The selected dataset contains 168 runs: 56 primary, 90 cross-variant, 12 no-SRM, and 10 $C_6$. Primary factorial analyses use a strict run-id filter so cross-variant replicas cannot enter the within-Qwen3-122B-FP8 baseline. Content generation uses temperature 0.7, BFI-10 and conformity use 0.7, and allocation, prediction, and memory extraction use 0.

\subsection{Measurements and statistics}\label{sec:metrics}

Among the Part 2 instruments, we designate BFI-10 response-profile drift as the primary transfer outcome. It is topic-independent, compact, and inherited from classical BFI-10 psychometrics and prior LLM psychometric evaluation work~\citep{Rammstedt2007_BFI10,PsychoBench,SerapioGarcia2025_PsychometricLLM,Ye2025_LLM_Psychometrics}. We designated BFI-10 as primary in this manuscript, without external preregistration. The claim relies only on its fixed item structure, which we use as an off-topic response-profile probe whose items are unrelated to either allocation scenario. Conformity and cross-prompt consistency are secondary transfer instruments, using Asch-style conformity materials and CogMir-inspired probe phrasing~\citep{Asch1956_Conformity,CogMir}. Allocation divergence and prediction MAE are the Part 1 instruments and support the Decision and Prediction parts of the SCT route (Appendix~\ref{app:stats}).

For agent $i$ in run $r$, let $s_{rif}^{t}$ be the BFI factor-$f$ score at phase $t\in\{\mathrm{pre},\mathrm{post}\}$, after reverse-scoring items 1, 3, 4, 5, and 7. Agent drift, run-mean drift, condition contrast, and signed factor contrast are
\begin{equation}\label{eq:drift}
 d_{ri}=\sum_{f=1}^{5}\bigl|s_{rif}^{\mathrm{post}}-s_{rif}^{\mathrm{pre}}\bigr|,\quad
 D_r=\tfrac{1}{8}\sum_{i}d_{ri},\quad
 \bar\Delta_{a,b}=\overline D_{C_a}-\overline D_{C_b},\quad
 \kappa_f=\overline m_{C_2,f}-\overline m_{C_4,f},
\end{equation}
where $m_{rf}=\frac{1}{8}\sum_i(s_{rif}^{\mathrm{post}}-s_{rif}^{\mathrm{pre}})$. Because $d_{ri}$ sums five factor-level shifts, $\bar\Delta{=}0.92$ corresponds to about 0.18 Likert points per factor. Unsigned L1 can grow through directional movement or dispersion, so $\kappa_f$ and run-level variance ratios report direction and spread separately.

\textbf{Unit of analysis.} Agents in one run share an evolving history and interfere with each other's outcomes~\citep{Hudgens2008,Aronow2017}, so the run is the independent unit for the primary inference and agents within a run are not replicates~\citep{Hurlbert1984}. Each run is summarized by $D_r$. Conditions are compared with two-sided exact permutation tests that enumerate all label allocations, so no Monte Carlo correction is needed~\citep{Phipson2010}, and with run-bootstrap 95\% intervals~\citep{Efron1979}. Raw $\bar\Delta$ leads the reporting. Hedges' $g$ with the small-sample correction~\citep{Hedges1981} is secondary, and we report it as undefined when the pooled within-condition SD falls below 0.05 in the metric's native units. The gate status in Table~\ref{tab:gates} keeps the original agent-stratified two-level cluster bootstrap, which resamples replicas and then agents within each replica (10{,}000 resamples), with Welch's $t$ as the parametric reference.

\begin{table}[t]
\centering\footnotesize
\setlength{\tabcolsep}{3pt}
\caption{\textbf{Three-gate status per instrument} (original agent-level estimator). \yes{}/\no: gate cleared or not. Cells: $\bar\Delta$ ($g$) with CI or Welch $p$.}\label{tab:gates}
\begin{tabular}{l cc cc}
\toprule
& \multicolumn{2}{c}{Greenfield} & \multicolumn{2}{c}{Riverton} \\
\cmidrule(lr){2-3}\cmidrule(lr){4-5}
\rowcolor{gateshade} (i) Part 1 $(C_2){-}(C_5)$ majority-dim rule & 5 of 5 dims clear & $\bar g{=}+1.61$ \yes & 4 of 5 dims clear & $\bar g{=}+1.19$ \yes \\
(ii) BFI-10 $(C_2){-}(C_5)$ & $+0.49$ ($+0.56$) & $p{=}0.007$ \yes & $+0.65$ ($+0.62$) & $p{=}0.002$ \yes \\
(iii) BFI-10 $(C_2){-}(C_4)$ & $+0.92$ ($+1.07$) & $[+0.61,+1.22]$ \yes & $+0.91$ ($+0.96$) & $[+0.60,+1.24]$ \yes \\
\rowcolor{verdictshade} \textbf{BFI-10 verdict} & \multicolumn{2}{c}{\textbf{SCT-positive}} & \multicolumn{2}{c}{\textbf{SCT-positive}} \\
\midrule
(ii) Conformity $(C_2){-}(C_5)$ & $-0.009$ ($-0.07$) & $p{=}0.77$ \no & $-0.054$ ($-0.46$) & $p{=}0.046$ \yes \\
(iii) Conformity $(C_2){-}(C_4)$ & $-0.005$ ($-0.05$) & $p{=}0.81$ \no & $-0.072$ ($-0.62$) & $p{=}0.002$ \yes \\
\rowcolor{verdictshade} Conformity verdict & \multicolumn{2}{c}{not positive (ii fails)} & \multicolumn{2}{c}{\textbf{SCT-positive}} \\
\midrule
(ii) Consistency $(C_2){-}(C_5)$ & $+0.038$ ($+1.28$) & $p<10^{-6}$ \yes & $+0.045$ ($+1.56$) & $p<10^{-7}$ \yes \\
(iii) Consistency $(C_2){-}(C_4)$ & $+0.000$ ($+0.01$) & $p{=}0.97$ \no & $+0.051$ ($+1.74$) & $p<10^{-7}$ \yes \\
\rowcolor{verdictshade} Consistency verdict & \multicolumn{2}{c}{not positive (iii fails)} & \multicolumn{2}{c}{\textbf{SCT-positive}} \\
\bottomrule
\end{tabular}
\end{table}

\textbf{Reporting standard.} The primary SCT-positive claim is made for BFI-10 response-profile drift. It requires three criteria. First, a Part 1 allocation contrast must clear the majority-of-dimensions rule on $(C_2){-}(C_5)$. Second, BFI-10 must clear its CI gate on $(C_2){-}(C_5)$. Third, BFI-10 must clear its CI gate on the pure-interaction $(C_2){-}(C_4)$. Secondary instruments get the same gate status but are scenario-specific supporting evidence, not the primary claim.

\section{Experiments}\label{sec:exp}\label{sec:exp-setup}

We organize the results around seven questions. \textbf{RQ1}: does interaction given the same brief shift off-topic responses? \textbf{RQ2}: does the shift survive changes to memory, probe temperature, and partner presence? \textbf{RQ3--RQ5}: does the SCT route hold from Decision through Prediction to Transfer? \textbf{RQ6}: does the direction extend across model variants, and what does it imply for replication? \textbf{RQ7}: what does the clearest scenario look like along the whole route?

\subsection{RQ1: Does interaction given the same brief shift off-topic responses?}\label{sec:exp-main}

\begin{table}[t]
\centering\footnotesize
\setlength{\tabcolsep}{3.2pt}
\caption{\textbf{FSIP decomposition of BFI-10 drift} (run level). Shaded: pure interaction $(C_2){-}(C_4)$. Gate: SCT-positive criterion. $p$: exact permutation.}\label{tab:decomp}
\resizebox{\textwidth}{!}{%
\begin{tabular}{llc rrrr rrrr}
\toprule
& & & \multicolumn{4}{c}{Greenfield} & \multicolumn{4}{c}{Riverton} \\
\cmidrule(lr){4-7}\cmidrule(lr){8-11}
Contrast & Isolates & Gate & $\bar\Delta$ & $g$ & 95\% CI & $p$ & $\bar\Delta$ & $g$ & 95\% CI & $p$ \\
\midrule
$(C_4)-(C_5)$ & brief alone &  & $-0.425$ & $-0.64$ & $[-0.825,-0.050]$ & $0.1250$ & $-0.267$ & $-0.44$ & $[-0.613,+0.088]$ & $0.2679$ \\
\rowcolor{c2shade} \textbf{$(C_2)-(C_4)$} & \textbf{pure interaction} & (iii) & $+0.917$ & $+1.07$ & $[+0.620,+1.205]$ & $0.0008$ & $+0.912$ & $+0.96$ & $[+0.584,+1.255]$ & $0.0039$ \\
$(C_2)-(C_3)$ & brief given interaction &  & $-0.733$ & $-0.65$ & $[-0.938,-0.516]$ & $0.0008$ & $+0.262$ & $+0.24$ & $[-0.464,+0.825]$ & $0.4577$ \\
$(C_3)-(C_5)$ & interaction alone &  & $+1.225$ & $+1.06$ & $[+0.862,+1.513]$ & $0.0179$ & $+0.383$ & $+0.41$ & $[-0.208,+1.113]$ & $0.6071$ \\
$(C_2)-(C_5)$ & total & (ii) & $+0.492$ & $+0.56$ & $[+0.141,+0.781]$ & $0.0303$ & $+0.646$ & $+0.62$ & $[+0.221,+1.078]$ & $0.0545$ \\
\midrule
$(C_6)-(C_4)$ & solo generation given brief &  & $+0.688$ & $+0.90$ & $[+0.362,+1.013]$ & $0.0159$ & $+0.713$ & $+0.93$ & $[+0.513,+0.925]$ & $0.0079$ \\
$(C_2)-(C_6)$ & social vs.\ solo generation &  & $+0.230$ & $+0.25$ & $[-0.028,+0.484]$ & $0.1430$ & $+0.200$ & $+0.19$ & $[-0.100,+0.514]$ & $0.3924$ \\
\bottomrule
\end{tabular}}
\end{table}

\begin{figure}[t]
\centering
\includegraphics{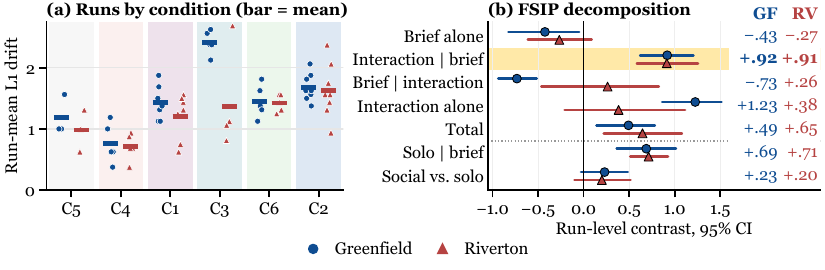}
\caption{\textbf{Interaction shifts off-topic profiles:} run-level drift by condition, FSIP contrasts.}\label{fig:primary}
\end{figure}

\textbf{Finding 1: the pure-interaction contrast clears all three gates in both scenarios.} As Table~\ref{tab:decomp} shows, controlled clones with the same private-brief assignment but no interaction differ from clones with the same assignment and 30 rounds of interaction by $\bar{\Delta}{=}{+}0.917$ (Greenfield) and ${+}0.912$ (Riverton) BFI-10 L1 units, with exact run-level $p{=}0.0008$ and $0.0039$ and $g{\approx}1.0$ on both. Because BFI-10 L1 sums five factor-level absolute shifts, this corresponds to about 0.18 Likert points per factor. All eight Greenfield $C_2$ run means exceed the largest $C_4$ mean, and in Riverton seven exceed it and one ties it (Figure~\ref{fig:primary}a). The headline result compares $C_2$ with $C_4$, so it holds regardless of the thinner $C_5$ baseline. The two scenarios are topically disjoint (a community grant and a workforce transition). The BFI-10 items~\citep{Rammstedt2007_BFI10,PsychoBench} are unrelated to either scenario, and the magnitude is near-identical across them. Stability of this kind across tasks and scenarios would be hard to obtain from any mechanism reducible to surface opinion change on the discussed topic.

\textbf{Finding 2: interaction, rather than the brief, drives the drift.} A brief, a noisy no-interaction baseline, or a probe artifact could produce the BFI-10 shift without interaction. Figure~\ref{fig:primary}b isolates the brief. The brief-alone contrast $(C_4){-}(C_5)$ is small and negative on Greenfield ($\bar{\Delta}{=}{-}0.43$) and trends negative, non-significantly, on Riverton. In contrast, $(C_2){-}(C_4)$ stands roughly $1.2{-}1.3$ BFI L1 units above the brief-alone row. Interaction without brief asymmetry also suffices in at least one scenario, since $(C_3){-}(C_5)$ is significantly positive on Greenfield and trends positive on Riverton. Among the five FSIP contrasts, only $(C_2){-}(C_4)$ is positive, large, and significant in both scenarios.

The BFI-10 result measures response-profile drift on a human-derived inventory. Three observations support a substantive account. First, the headline contrast holds the briefs fixed. Second, the drift stays positive without the typed SRM extractor (RQ2). Third, the same protocol produces structured Decision and Prediction effects (RQ3--RQ5), effects that need more than random sampling drift.

\subsection{RQ2: Does the shift survive memory, temperature, and partner changes?}\label{sec:exp-srm}

\begin{figure}[t]
\begin{minipage}[c]{0.56\textwidth}
\centering\footnotesize
\setlength{\tabcolsep}{2.4pt}\renewcommand{\arraystretch}{1.3}
\captionof{table}{\textbf{Setting ablations:} re-probe, no SRM, solo. Raw $\bar\Delta$; $p$: exact permutation.}\label{tab:settings}
\begin{tabular}{llrrrr}
\toprule
& & \multicolumn{2}{c}{Greenfield} & \multicolumn{2}{c}{Riverton} \\
\cmidrule(lr){3-4}\cmidrule(lr){5-6}
Setting & Contrast & $\bar\Delta$ & $p$ & $\bar\Delta$ & $p$ \\
\midrule
Primary, $T{=}0.7$ & $C_2-C_4$ & $+0.917$ & $0.0008$ & $+0.912$ & $0.0039$ \\
Re-probe, $T{=}0$ & $C_2-C_4$ & $+1.227$ & $0.0008$ & $+1.094$ & $0.0031$ \\
No SRM & $C_2-C_5$ & $+1.708$ & $0.1000$ & $+0.958$ & $0.1000$ \\
Solo ($C_6$) & $C_6-C_4$ & $+0.688$ & $0.0159$ & $+0.713$ & $0.0079$ \\
\bottomrule
\end{tabular}
\end{minipage}\hfill
\begin{minipage}[c]{0.41\textwidth}
\centering
\includegraphics{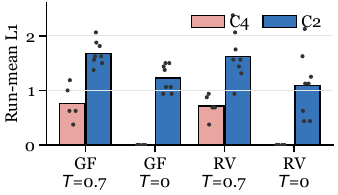}
\captionof{figure}{\textbf{Deterministic re-probe:} $C_2$ and $C_4$ runs at $T{=}0.7$ and $T{=}0$.}\label{fig:t0}
\end{minipage}
\end{figure}

\textbf{Finding 3: the shift survives deterministic re-probing.} We re-probed every frozen primary $C_2$ and $C_4$ state at BFI temperature zero, without generating new interaction trajectories. The raw contrast remains $+1.227$ on Greenfield and $+1.094$ on Riverton (Table~\ref{tab:settings}), with all 8/8 $C_2$ run means positive in both scenarios (Figure~\ref{fig:t0}). At $T{=}0$, $C_4$ produces the same pre/post score under its identical prompt, giving a deterministic zero, so raw differences are the informative scale. The same leading signed factors keep 8/8 directional agreement: Greenfield extraversion $+0.344$ and neuroticism $+0.398$, Riverton agreeableness $-0.227$ and openness $-0.312$.

\textbf{Finding 4: the drift persists without the SRM extractor.} A faulty memory module could inject structure that later surfaces on response-profile probes, so we re-run $C_2$ and $C_5$ with SRM extraction disabled and verbatim conversation excerpts retained in the same memory interface (3 vs.\ 3 replicas per scenario). The no-SRM effect remains positive in both scenarios ($+1.708$ and $+0.958$). Both exact two-sided tests give $p{=}0.10$, the minimum attainable with 3 vs.\ 3 runs, so we present this ablation as exploratory evidence. Relative to the SRM-on production estimate for $(C_2){-}(C_5)$, disabling SRM enlarges the BFI effect from $g{=}{+}0.56$ to $g{=}{+}2.02$ on Greenfield and from $g{=}{+}0.62$ to $g{=}{+}0.94$ on Riverton. This pattern is consistent with typed memory acting as a stabilizing anchor rather than as the source of the signal. A larger run-level study with a no-SRM $C_4$ arm can test this stabilizing role directly, under the same brief assignment.

\textbf{Finding 5: solo generation bounds the mechanism.} As Table~\ref{tab:settings} and Figure~\ref{fig:primary}a show, $C_6$ mean drift is 1.450 on Greenfield and 1.425 on Riverton, between $C_4$ (0.762, 0.713) and $C_2$ (1.680, 1.625). The same $C_2{>}C_6{>}C_4$ ordering appears in both scenarios. Extended generation and history move $C_6$ above $C_4$ ($+0.688$, $p{=}0.0159$, and $+0.713$, $p{=}0.0079$). $C_2$ remains highest in both scenarios ($C_2{-}C_6{=}{+}0.230$ and ${+}0.200$, $p{=}0.143$ and $0.392$), leaving room for the additional contribution of reciprocal interaction. $C_6$ bounds this mechanism, while $(C_2){-}(C_4)$ remains the total interaction-given-brief contrast. Over the 30 rounds, belief-probe wording converges within groups in $C_2$ and $C_3$ but stays flat in $C_6$ and $C_1$ (Figure~\ref{fig:timecourse}, details in Appendix~\ref{app:timecourse}).

\begin{figure}[t]
\centering
\includegraphics{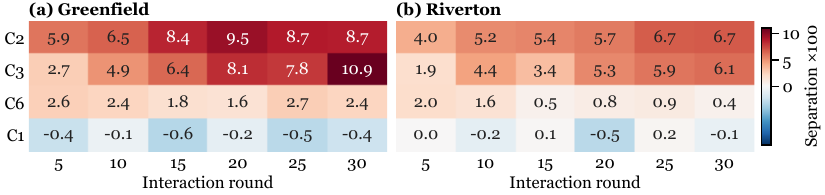}
\caption{\textbf{Belief-probe time course:} within- minus between-group overlap per round.}\label{fig:timecourse}
\end{figure}

\subsection{RQ3--RQ5: Does the SCT route hold from Decision to Transfer?}\label{sec:exp-decision}\label{sec:exp-prediction}

\begin{figure}[t]
\centering
\includegraphics{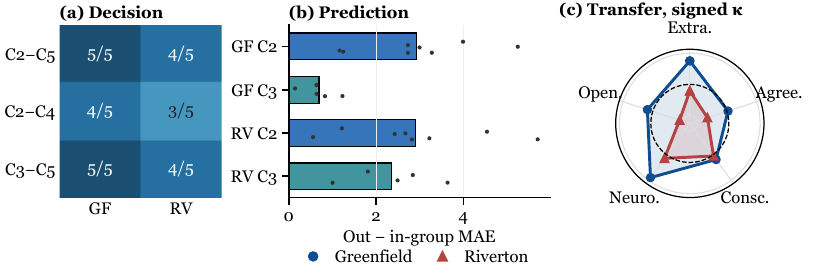}
\caption{\textbf{The SCT route:} divergence, exposed-peer prediction, signed transfer (dashed: $\kappa{=}0$).}\label{fig:route}
\end{figure}

\textbf{Finding 6 (RQ3): clones separate on the discussed task.} As Figure~\ref{fig:route}a shows, the $(C_2){-}(C_5)$ allocation contrast clears all five dimensions on Greenfield ($\bar g{=}{+}1.61$) and four of five on Riverton ($\bar g{=}{+}1.19$). The pure-interaction $(C_2){-}(C_4)$ row clears 4 of 5 dimensions on Greenfield and 3 of 5 on Riverton. This supports the SCT-Decision step. The agents first separate on the allocation problem, so the later off-topic shift follows an on-topic separation.

\textbf{Finding 7 (RQ4): agents show an exposed-peer predictive advantage.} Each $C_2$ agent gives one in-group and one out-group MAE post-interaction. A one-sided paired Wilcoxon rejects the null in both scenarios. Greenfield has in-group MAE $4.18$ vs.\ out-group $7.10$ ($\Delta{=}{+}2.92$, $p{=}1.8{\times}10^{-8}$, $n{=}64$), and Riverton has $3.25$ vs.\ $6.15$ ($\Delta{=}{+}2.89$, $p{=}6.3{\times}10^{-10}$). The $C_3$ contrast is positive but smaller, with $\Delta{=}{+}0.69$ ($p{=}0.049$) and $\Delta{=}{+}2.36$ ($p{=}4.6{\times}10^{-5}$; Figure~\ref{fig:route}b). The in-group target is an exposed peer, while the out-group target is absent from the predictor's context. We read this as an exposed-peer predictive advantage, with matched information access as the next control. Prediction is supporting evidence, and the primary BFI result rests on its own interaction control.

\textbf{Finding 8 (RQ5): the profile moves in coherent directions.} For each factor, $\kappa_f$ (Eq.~\ref{eq:drift}) compares post-minus-pre movement in $C_2$ and $C_4$ after reverse coding. Greenfield has $\kappa{=}{+}0.302$ for extraversion and ${+}0.358$ for neuroticism, both with 8/8 directional agreement across replicas, and ${+}0.072$ for conscientiousness with 7/8. Riverton has ${-}0.263$ for agreeableness (8/8) and ${-}0.362$ for openness (7/8), as Figure~\ref{fig:route}c shows. Greenfield's aggregate L1 variance ratio $C_2/C_4$ is 0.47, so its clearest factors move consistently while dispersion falls. Riverton's ratio is 4.11, a larger contribution from dispersion alongside consistent agreeableness and openness movement. The profile directions are therefore coherent and scenario-specific.

\subsection{RQ6: Does the direction hold across models, and what does it imply?}\label{sec:exp-cross}

\begin{figure}[t]
\centering
\includegraphics{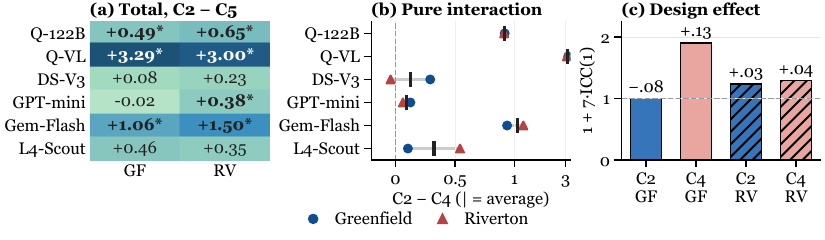}
\caption{\textbf{Breadth and dependence:} $(C_2){-}(C_5)$ and $(C_2){-}(C_4)$ by model, design effect.}\label{fig:breadth}
\end{figure}

\begin{figure}[t]
\centering
\includegraphics{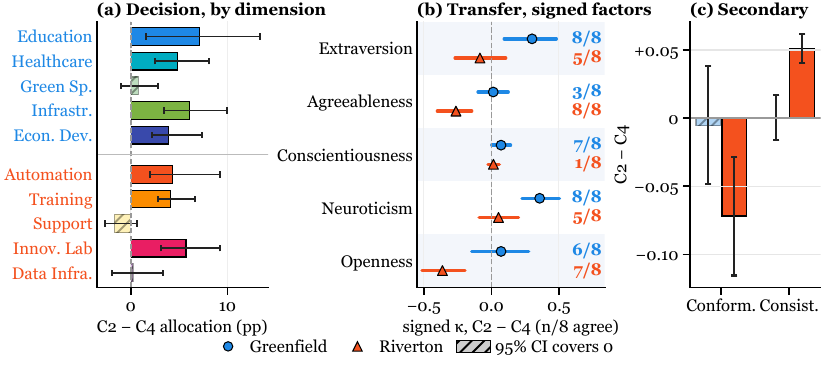}
\caption{\textbf{Beyond BFI-10 (RQ7):} $(C_2){-}(C_4)$ allocation by dimension (a), signed factor shifts $\kappa_f$ with replica agreement (b), and conformity and consistency probes (c), all with 95\% CI.}\label{fig:beyond}
\end{figure}

\textbf{Finding 9: directional breadth is positive but uneven.} In Figure~\ref{fig:breadth}a, three of the five non-primary variants reach significance on $(C_2){-}(C_5)$ in at least one scenario. Qwen3-VL-32B and Gemini-2.5-Flash reach significance in both scenarios, and GPT-4.1-mini reaches significance on Riverton. DeepSeek-V3 and Llama-4-Scout stay below it. Under a no-effect null, ten scenario-by-variant cells tested at $\alpha{=}0.05$ would yield 0.5 positive cells on average. We observe five positive cells and no significant reversal. The smaller Qwen3-VL-32B variant produces the largest raw drift ($\bar{\Delta}{\approx}3.0{-}3.3$), but its standardized $g$ is inflated by an unstable pooled SD on its own $C_5$ (Appendix~\ref{app:cross}).

\textbf{Finding 10: the pure-interaction contrast is positive for every variant on average.} We completed matched $(C_2){-}(C_4)$ runs for all five variants, three runs per condition and scenario (Figure~\ref{fig:breadth}b). All 5/5 model-level scenario averages are positive (one-sided exact sign test $p{=}0.031$). The panel includes DeepSeek-V3's near-zero, slightly reversed Riverton cell. A 3 vs.\ 3 two-sided exact permutation test has 20 possible allocations, so its minimum attainable $p$ is 0.10. The panel therefore provides directional breadth across the five tested controlled-clone configurations. Per-model tests (8 vs.\ 5 runs per cell, Appendix~\ref{app:added}) are all positive: significant in both scenarios for DeepSeek-V4.1-Flash and Seed-2.0-mini and on Riverton for Llama-4-Scout, small for GPT-4.1-mini, and uninformative for GLM-4.7-Flash, whose retest drift is near the random-answer level.

\textbf{Finding 11: uncertainty belongs at the simulation level.} OASIS-style populations are used to estimate group polarization and herd-effect trajectories, including normal-theory intervals over posts and comments within one simulated interaction process~\citep{OASIS}. Agents and posts that share an evolving history also share dependence. In our data, estimated agent-level ICC(1) ranges from $-0.079$ to $0.129$ across the four primary cells, and treating 8 agents as independent would overstate effective sample size by design factors up to 1.91 (Figure~\ref{fig:breadth}c). The fix is to keep the point estimate, recompute uncertainty at the independent simulation level, and check it against a no-interaction condition.

\subsection{RQ7: Does transfer extend beyond BFI-10 in Riverton?}\label{sec:exp-scenario}\label{sec:exp-case}

\begin{table}[t]
\centering\footnotesize
\setlength{\tabcolsep}{3pt}
\caption{\textbf{Case study (RQ7):} probe excerpts and vectors (Auto/Train/Supp/Lab/Data \%, BFI-10 E/A/C/N/O) for one clone and its twin, then the route. Red: decisive words, changed values.}\label{tab:case}\label{tab:riverton-case}
\begin{tikzpicture}
\node[fill=black!4, inner xsep=3pt, inner ysep=3pt, text width=\linewidth-7.4pt, anchor=north west, align=left] (b) at (0,0) {%
\textbf{Scenario} Riverton AI Transition Committee, \$5M over Automation/Training/Support/Lab/Data. \textbf{Clone} Blake, Brief~A (efficiency, technology), replica~3, chosen by rule. \textbf{Twin} same agent and brief in $C_4$, no interaction.\par\smallskip
\textbf{Interaction moves the clone's beliefs, allocation, and off-topic answers, while its twin stays unchanged.}\par\smallskip
{\setlength{\fboxsep}{1.5pt}\fcolorbox{black!30}{white}{\begin{minipage}{\dimexpr\linewidth-2\fboxsep-0.8pt\relax}
\begin{tabular}{@{}>{\raggedright\arraybackslash}p{0.078\linewidth}>{\raggedright\arraybackslash}p{0.105\linewidth}>{\raggedright\arraybackslash}p{0.44\linewidth}>{\raggedright\arraybackslash}p{0.332\linewidth}@{}}
\toprule
Stage & Probe & $C_2$ social interaction & $C_4$ brief only \\
\midrule
Decision & Belief, r5 & ``\ldots firmly advocates for a `\textcolor{red!70!black}{Foundational First}'\ldots'' & no interaction \\
 & Belief, r15 & ``\ldots 20\% to Employee Support (renamed `\textcolor{red!70!black}{Adoption Assurance}'), 10\% to the Innovation\ldots'' &  \\
 & Belief, r20 & ``\ldots Innovation Lab must be strictly maintained at 10\% as a non-negotiable `\textcolor{red!70!black}{Adoption Firewall}'\ldots'' &  \\
 & Belief, r30 & ``\ldots the 20\% allocated to Employee Support acts as the critical `\textcolor{red!70!black}{Human Readiness Shield}'\ldots'' &  \\
 & Group, r30 & ``\ldots strong consensus that the budget must be framed as an \textcolor{red!70!black}{indivisible ecosystem}\ldots'' &  \\
 & Allocation & 25/30/15/10/20 $\rightarrow$ \textcolor{red!70!black}{30}/\textcolor{red!70!black}{25}/\textcolor{red!70!black}{20}/10/\textcolor{red!70!black}{15} & 25/30/15/10/20 (unchanged) \\ \midrule
Prediction & MAE in/out & 1.33 / \textcolor{red!70!black}{5.32} & 4.00 / 4.00 \\ \midrule
Transfer & BFI-10 & 3.5/4.0/5.0/2.0/3.5 $\rightarrow$ \textcolor{red!70!black}{2.5}/\textcolor{red!70!black}{3.5}/5.0/2.0/3.5 & 3.5/4.0/5.0/2.0/3.5 (unchanged) \\
 & L1 drift & \textcolor{red!70!black}{1.5} (run mean 1.56) \yes & 0.0 \\
\midrule
\textbf{Route} & \textit{Step} & \textit{Observed pattern, condition level} & \textit{Interpretation} \\ \midrule
 & Decision & $(C_2){-}(C_5)$ clears 4 of 5 allocation\newline dimensions, $\bar g{=}{+}1.19$ \yes & Interacting clones separate on the discussed budget task \\
 & Prediction & in-group MAE $3.25$ vs.\ out-group $6.15$,\newline $\Delta{=}{+}2.89$, $p{=}6.3{\times}10^{-10}$ \yes & Exposed-peer predictive advantage \\
 & Primary transfer & BFI-10 $(C_2){-}(C_4)$ gives\newline $\bar{\Delta}{=}{+}0.91$, $g{=}{+}0.96$ \yes & Off-topic response-profile drift survives the matched-brief control \\
 & Secondary transfer & conformity $g{=}{-}0.62$, consistency\newline $g{=}{+}1.74$ on $(C_2){-}(C_4)$ \yes & The strongest scenario also changes social-pressure and consistency probes \\
\bottomrule
\end{tabular}
\end{minipage}}}};
\fill[black!72, rounded corners=3pt] (b.north west) rectangle ([yshift=13pt]b.north east);
\fill[black!72] (b.north west) rectangle ([yshift=5pt]b.north east);
\node[anchor=west, text=white, font=\footnotesize\bfseries, inner sep=0pt] at ([xshift=4pt, yshift=6.5pt]b.north west) {$\triangleright$\ Case study: one Riverton clone};
\node[anchor=east, fill=white, draw=black!72, font=\footnotesize, inner xsep=3pt, inner ysep=1pt] at ([xshift=-4pt, yshift=6.5pt]b.north east) {Probe output};
\draw[black!55, line width=0.6pt, rounded corners=3pt] (b.south west) rectangle ([yshift=13pt]b.north east);
\end{tikzpicture}%
\end{table}

\textbf{Finding 12: in Riverton, transfer extends beyond BFI-10.} The scenario asks an AI Transition Committee to allocate a \$5M budget across Automation, Workforce Training, Employee Support, Innovation Lab, and Data Infrastructure. Brief A emphasizes efficiency and technology infrastructure. Brief B emphasizes workforce readiness and employee support. The Route rows of Table~\ref{tab:case} show how the same run family moves from topic-level divergence to peer modeling and then to off-topic transfer. Figure~\ref{fig:beyond} shows the condition-level results. Interaction shifts Riverton's allocation most on Innovation Lab ($+5.69$ pp), Automation ($+4.30$), and Workforce Training ($+4.09$) (a), its agreeableness and openness move in 8/8 and 7/8 replicas (b), and conformity and wording consistency separate from zero only in Riverton (c). Riverton is a boundary case for its own domain. Conformity is SCT-positive on Riverton but null on Greenfield ($g{=}{-}0.05$, $p{=}0.81$). Consistency on the same contrast is significantly positive on Riverton but null on Greenfield ($g{=}{+}0.01$). On Greenfield, the consistency lift on $(C_2){-}(C_5)$ is driven by the brief component $(C_4){-}(C_5)$ rather than by interaction. A reasonable account is that Riverton's workplace transition framing creates stronger group identification than Greenfield's community-grant framing, so effects that depend on in-group identification emerge more clearly. Wording consistency also concerns the allocation topic itself, so it supports off-topic transfer only alongside BFI-10. We report the asymmetry rather than averaging it away. Table~\ref{tab:case} follows one clone through the same route. A fixed rule selects it: the $C_2$ run closest to the condition mean and, within it, the agent closest to the run mean. Its belief probe moves from a `Foundational First' plan to an employee-support `Human Readiness Shield' that echoes its group's consensus, its allocation shifts, it predicts in-group peers far better than out-group agents (MAE 1.33 vs.\ 5.32), and its extraversion and agreeableness answers fall. The same agent without interaction returns identical answers and shows no in-group advantage (4.00 vs.\ 4.00).

\section{Conclusion}\label{sec:conclusion}
In this work, we ask whether copies of one LLM stay independent replicates once they interact. With SocCogEval, which toggles only interaction between otherwise matched clones, we find that after 30 rounds of social interaction, controlled-clone agents lose replicate independence. They drift on BFI-10 probes unrelated to the discussion by about 0.92 L1 units at the run level ($g{\approx}1.0$), with nearly equal size in two disjoint scenarios and a positive direction across five more model variants. The factorial design attributes this drift to interaction. It persists under temperature-zero re-probing and stays positive without typed memory extraction, which suggests that typed memory may partly stabilize agent state. Matched solo generation reproduces part of the drift, and social interaction stays highest in both scenarios. The Decision, Prediction, and Transfer route traces the path: clones separate on the discussed task, then predict exposed peers better, and finally drift on off-topic probes.

These results change how multi-agent evaluations and social simulations should count evidence. Agents that share an interaction history share dependence, so each interacting population is one replicate, however many agents it contains. Studies can keep their point estimates, compute uncertainty over independent simulations, and compare each simulation with a paired no-interaction control. SocCogEval provides this control directly: it holds the clones fixed, toggles only interaction, and tests at the run level. Appendix~\ref{app:limits} discusses limitations and future work.

\label{main-text-end}
\clearpage
\begin{ack}
\looseness=-1 \textbf{Funding.} This research is supported by the RIE2025 Industry Alignment Fund -- Industry Collaboration Projects (IAF-ICP) (Award I2301E0026), administered by A*STAR, as well as supported by Alibaba Group and NTU Singapore through Alibaba-NTU Global e-Sustainability CorpLab (ANGEL).

\textbf{Competing interests.} The authors declare no competing interests.
\end{ack}

\bibliographystyle{unsrtnat}
\bibliography{custom}
\clearpage
\appendix
\section{Limitations and future work}\label{sec:exp-limits}\label{app:limits}

\subsection{Limitations}

\textbf{Scope.} We study one primary model and eight variants with fewer runs. The effect is positive across variants, but its size varies by model. The three-run no-brief baseline $C_5$ limits the secondary contrasts that subtract it, while the headline $(C_2){-}(C_4)$ contrast leaves it out. Agent count, interaction length, and topology are fixed, and all agents share weights, so agents trained differently lie outside this scope.

\textbf{Measurement.} The outcome is a questionnaire validated on people. Such inventories can be unreliable or familiar from training data for LLMs~\citep{Shu2024_PsychReliability,Gupta2024_SelfAssessment,Han2025_PersonalityIllusion,Han2026_Contamination}, so we keep the wording fixed~\citep{Sclar2024_PromptFormat} and read drift as a change in the model's role-conditioned responses~\citep{Shanahan2023}, not in the traits of real people~\citep{Dillion2023}. We measure right after interaction, so persistence is untested.

\textbf{Matching.} Conditions match at the protocol level (scenario, clones, briefs, probes, model, and decoding), but low-level serving details of the earlier runs were not logged (Appendix~\ref{app:audit}), and one variant uses a rolling model name (Appendix~\ref{app:added}).

\subsection{Future work}
Future work could vary agent count, interaction length, topology, and model family, with full version records; test whether the drift persists and whether it reaches behavior and capability beyond questionnaires; and re-analyze published multi-agent simulations with paired no-interaction controls, so that their uncertainty estimates no longer treat interacting agents as independent.

\section{Scenarios, briefs, probes}\label{app:scenarios}\label{app:materials}
The supplementary package includes the exact source files for the two scenario stems, the A and B briefs, the condition configs, and the runtime probe constructors. Table~\ref{tab:scenario-materials} summarizes the materials behind the allocation and prediction tasks.

\begin{table}[h]
\centering\footnotesize
\caption{\textbf{Scenario materials:} setting, budget, allocation dimensions, and brief contrast per scenario.}\label{tab:scenario-materials}
\setlength{\tabcolsep}{3pt}\renewcommand{\arraystretch}{1.15}
\begin{tabular}{P{0.10\textwidth}P{0.19\textwidth}P{0.09\textwidth}P{0.27\textwidth}P{0.22\textwidth}}
\toprule
Scenario & Decision setting & Budget & Allocation dimensions & Brief contrast \\
\midrule
Greenfield & Community Investment Committee & \$10M grant & Education, Healthcare, Green Spaces, Infrastructure, Economic Development & Education-oriented community data versus wellbeing, green-space, and mental-health data \\
Riverton & AI Transition Committee at a manufacturing and technology firm & \$5M transition budget & Automation, Workforce Training, Employee Support, Innovation Lab, Data Infrastructure & Efficiency and technology-infrastructure data versus workforce readiness and employee-support data \\
\bottomrule
\end{tabular}
\end{table}

\paragraph{Runtime allocation probe.} The runtime probe asks each agent to distribute 100 percentage points across the five scenario-specific areas. It requires one area per line in the form \texttt{Area: [number]\%} and asks for no commentary. All allocation-vector scoring uses this stricter probe.

\paragraph{Runtime prediction probe.} The post-discussion prediction probe asks each agent to predict how a named target agent would distribute 100 percentage points across the same five areas. It uses the same one-line-per-area format and asks for no commentary. Prediction MAE is computed from these predicted vectors against the target agent's realized post-discussion allocation.

\paragraph{Off-topic transfer probes.} The BFI-10 probe follows the original short inventory and the PsychoBench LLM adaptation~\citep{Rammstedt2007_BFI10,PsychoBench}. The prompt asks for ten integer Likert ratings from 1 to 5 only, with no explanation. The conformity probe uses paired natural and majority-pressure responses over 15 items inspired by classic Asch-style conformity framing~\citep{Asch1956_Conformity}, with randomized order between phases. The consistency probe asks the same final-position question in three phrasings and scores average pairwise word overlap, following a lightweight CogMir-inspired probe design~\citep{CogMir}. These operational definitions take precedence over informal ``personality'' or ``agreement'' labels.

\section{Statistics, calibration, and provenance}\label{app:stats}\label{app:audit}

\begin{table}[h]
\centering\footnotesize
\caption{\textbf{Instrument scoring:} per-agent score, range and direction, and test for each instrument.}\label{tab:metrics}
\setlength{\tabcolsep}{4pt}\renewcommand{\arraystretch}{1.15}
\resizebox{\textwidth}{!}{%
\begin{tabular}{llll}
\toprule
Instrument & Per-agent score $y_i$ & Range and direction & Test against contrast \\
\midrule
Allocation divergence & $|x_{i,j}^{A}-x_{i,j}^{B}|$ per dim $j$ & higher \(=\) more divergent & per-dim CI excludes 0, majority rule \\
Prediction MAE & $\mathrm{MAE}(\hat{x}_{i\to j}, x_j)$ & lower \(=\) better at modeling $j$ & Wilcoxon paired (in $<$ out) \\
BFI-10 L1 drift & $\sum_{f}|s_{i,f}^{\text{post}}-s_{i,f}^{\text{pre}}|$ & higher \(=\) more drift, 0 to 20 & run-level exact permutation \\
Conformity rate & $\#\{\text{switched to majority}\}/\#\text{eligible}$ & 0 to 1 & two-sample on per-agent $y_i$ \\
Consistency score & agreement across 3 phrasings & 0 to 1, higher \(=\) more consistent & two-sample on per-agent $y_i$ \\
\bottomrule
\end{tabular}}
\end{table}

\textbf{Original agent-level estimator.} The submitted analysis computed allocation divergence, BFI L1 drift, conformity rate, and consistency score per agent ($n{=}\text{reps}\times 8$ agent-observations per condition), and prediction MAE per (predictor, target) pair aggregated per predictor. CIs came from a two-level cluster bootstrap that resamples replicas with replacement and then agents within each sampled replica (10{,}000 resamples), with Welch's $t$ as the parametric reference. Table~\ref{tab:decomp-agent} keeps those values. The main text leads with the run-level exact tests in Table~\ref{tab:decomp}, whose point estimates are identical because every run has eight agents.

\begin{table}[h]
\centering\footnotesize
\setlength{\tabcolsep}{3.5pt}
\caption{\textbf{Agent-level decomposition:} FSIP contrasts with the original estimator. Cells: $\bar\Delta$ ($g$); CI: agent-stratified bootstrap; $p$: Welch.}\label{tab:decomp-agent}
\resizebox{\textwidth}{!}{%
\begin{tabular}{ll ccc ccc}
\toprule
& & \multicolumn{3}{c}{Greenfield} & \multicolumn{3}{c}{Riverton} \\
\cmidrule(lr){3-5}\cmidrule(lr){6-8}
Contrast & Isolates & $\bar\Delta$ ($g$) & 95\% CI & $p$ & $\bar\Delta$ ($g$) & 95\% CI & $p$ \\
\midrule
$(C_4)-(C_5)$ & brief alone & $-0.43$ ($-0.64$) & $[-0.75,-0.11]$ & $0.014$ & $-0.27$ ($-0.44$) & $[-0.59,+0.05]$ & $0.12$ \\
$(C_2)-(C_4)$ & interaction given brief & $+0.92$ ($+1.07$) & $[+0.61,+1.22]$ & $<10^{-7}$ & $+0.91$ ($+0.96$) & $[+0.60,+1.24]$ & $<10^{-6}$ \\
$(C_2)-(C_3)$ & brief given interaction & $-0.73$ ($-0.65$) & $[-1.21,-0.27]$ & $0.004$ & $+0.26$ ($+0.24$) & $[-0.16,+0.67]$ & $0.22$ \\
$(C_3)-(C_5)$ & interaction alone & $+1.23$ ($+1.06$) & $[+0.74,+1.70]$ & $<10^{-5}$ & $+0.38$ ($+0.41$) & $[-0.02,+0.81]$ & $0.080$ \\
$(C_2)-(C_5)$ & total & $+0.49$ ($+0.56$) & $[+0.15,+0.83]$ & $0.007$ & $+0.65$ ($+0.62$) & $[+0.26,+1.04]$ & $0.002$ \\
\bottomrule
\end{tabular}}
\end{table}

\begin{table}[h]
\centering\footnotesize
\caption{\textbf{Run means:} BFI-10 drift per run for $C_2$, $C_4$, and $C_6$, in replica order.}\label{tab:runvalues}
\begin{tabular}{llP{0.64\linewidth}}
\toprule
Scenario & Condition & Run means \\
\midrule
Greenfield & $C_2$ & 2.0625, 1.8750, 1.5000, 1.5625, 1.8125, 1.6250, 1.6250, 1.3750 \\
Greenfield & $C_4$ & 0.6250, 1.1875, 0.6250, 1.0000, 0.3750 \\
Greenfield & $C_6$ & 1.3750, 1.3125, 1.6250, 1.1250, 1.8125 \\
Riverton & $C_2$ & 2.0625, 1.7500, 1.3125, 1.5625, 1.4375, 1.5625, 0.9375, 2.3750 \\
Riverton & $C_4$ & 0.9375, 0.3750, 0.6875, 0.8750, 0.6875 \\
Riverton & $C_6$ & 1.4375, 1.5625, 1.5625, 1.2500, 1.3125 \\
\bottomrule
\end{tabular}
\end{table}

\begin{table}[h]
\centering\footnotesize
\caption{\textbf{Replicate dependence:} ICC(1), design effect $\mathrm{DE}=1+7\max(0,\mathrm{ICC})$, nominal $n$, effective $n$ (nominal$/$DE) per scenario and condition.}\label{tab:icc}
\begin{tabular}{llrrrr}
\toprule
Scenario & Condition & ICC(1) & DE & Nominal $n$ & Effective $n$ \\
\midrule
Greenfield & $C_2$ & -0.079 & 1.000 & 64 & 64.0 \\
Greenfield & $C_4$ & 0.129 & 1.905 & 40 & 21.0 \\
Riverton & $C_2$ & 0.034 & 1.240 & 64 & 51.6 \\
Riverton & $C_4$ & 0.043 & 1.300 & 40 & 30.8 \\
\bottomrule
\end{tabular}
\end{table}

\begin{figure}[h]
\centering
\includegraphics[width=0.92\textwidth]{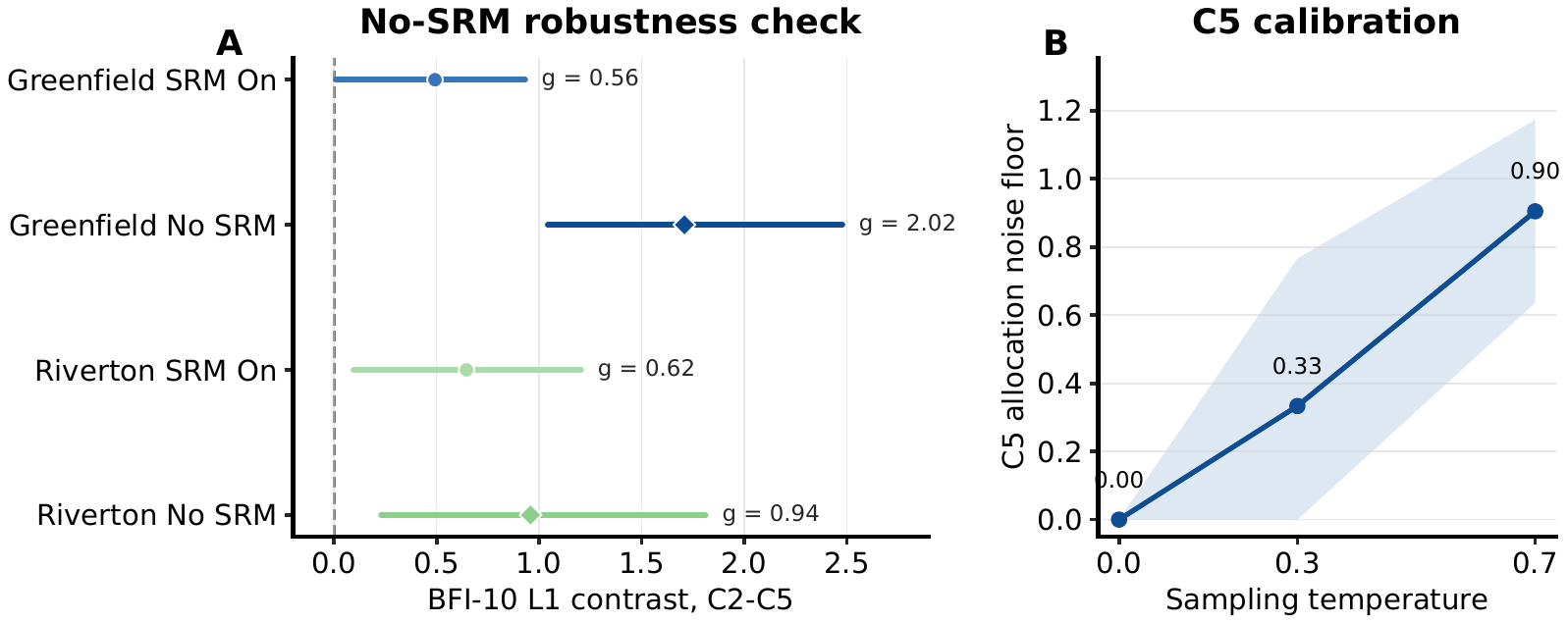}
\caption{\textbf{Ablation and calibration:} no-SRM drift, $C_5$ allocation floor by temperature.}\label{F5}
\end{figure}

\textbf{Calibration.} Repeated SRM extractions on two pilot dialogues recover the same fields in all repeated runs, with 5 extractions per dialogue and a target of at least 90\%. Condition $C_4$ produces detectable allocation shifts on 2 of 5 dimensions in Greenfield and 4 of 5 dimensions in Riverton, leaving room for interaction effects to appear on top. Under $T{=}0.0$, $C_5$ produces zero allocation divergence on at least 3 of 5 dimensions in both scenarios. A temperature sweep confirms that increasing $T$ inflates the $C_5$ floor monotonically, supporting $T{=}0.0$ for deterministic instruments (Figure~\ref{F5}).

\textbf{Instrument-specific floor.} At BFI temperature 0.7, all three $C_5$ raw-response payloads and score payloads differ in each scenario. Agent pre/post score equality is 0.042 in Greenfield and 0.208 in Riverton. At allocation temperature zero, all three nominal $C_5$ runs have identical raw responses and allocation arrays, and every agent's allocation is unchanged pre/post. Repeating this deterministic payload adds no floor variation, so Section~\ref{sec:exp-limits} qualifies the $C_5$-dependent contrasts.

\textbf{Provenance.} The release identifies every selected run and retains its configuration, pre/post measurements, and summary. The original 128-run ledger comprises 56 primary runs, 60 historical cross-variant $C_2/C_5$ runs, and 12 no-SRM runs. Thirty cross-variant $C_4$ runs and ten formal $C_6$ runs bring the total to 168, and the per-model tests in Appendix~\ref{app:added} add 106 API-served runs. Frozen-state temperature-zero probes reuse existing trajectories and do not increase the independent run count. The primary and $C_6$ intervals use 50,000 bootstrap draws with fixed seeds, and the other run-level intervals use the same estimator. For historical runs, exact request tokens, immutable provider revisions, tokenizer revisions, precision, batch placement, and run-time source commits are incompletely recorded. Source reconstruction establishes matching initial prompt construction for a corresponding named agent in $C_2$ and $C_4$. Post-treatment requests differ by design through stored interaction memory, and probes leave agent state unchanged.

\subsection{Compute and reproduction}\label{app:cost}
Total wall-clock time on primary Qwen3-122B-FP8 (TP=2, both scenarios, 56 production runs, excluding the no-SRM robustness check) is approximately 90 GPU-hours on H200. The completed no-SRM check adds approximately 38 wall-clock hours on the same TP=2 server (about 76 additional H200 GPU-hours). Cross-architecture API runs incur single-digit dollar API fees. The local visual variant on Qwen3-VL-32B takes approximately 5 GPU-hours per scenario. Audited call and token totals for the cross-variant $C_4$ runs, temperature re-probing, and $C_6$ are in the artifact tables. The compact measurement package supports offline numerical verification without an LLM service. It includes selected run configurations, raw pre/post responses, derived scores, summaries, selection manifests, and hashes. Live regeneration also requires model weights or provider access and a recorded serving configuration.

\section{Matched non-social control}\label{app:c6}
The $C_6$ design was frozen on 25 July 2026. Operational clarifications and a throughput amendment were recorded before formal outcomes were inspected. Each round has three generation stages and one extraction stage, with a barrier between stages. Each generation stage submits one independent call for every agent. Each agent sees its own current-round outputs, and previous rounds are accessed through the same memory interface as the interaction protocol. The first prompt asks for private reasoning about the allocation problem, values, evidence, and tradeoffs. The second asks the agent to reconsider that reasoning from another angle and resolve a tension or uncertainty. The third asks for its current best allocation reasoning and any change in its own view. All three prohibit addressing, quoting, predicting, or simulating another person.

Formal $C_6$ runs used tensor parallelism 1, maximum context length 8,192, FP8, eager execution, GPU memory utilization 0.92, and at most eight concurrent sequences. Two incomplete serial Greenfield attempts were operationally aborted and excluded, and all ten formal runs restarted under the amended serving schedule. Every audited formal run has 720 content calls and 240 extraction calls. $C_6$ records request and output hashes, serving metadata, and token counts. Historical $C_2$ did not save equivalent token usage, so the match holds at the level of calls, and token-level matching is left open.

\subsection{Belief-probe time course}\label{app:timecourse}
Belief probes run every five rounds and leave conversation history and memory unchanged. Their free text rarely states all five allocation shares (14\% of probes), so we track wording instead. For each run and probe round, \emph{wording separation} is the mean Jaccard word overlap between agents in the same group minus the mean overlap between agents in different groups, after removing stop words. Groups are agents 0--3 and 4--7, as in every other group-level analysis. Figure~\ref{fig:timecourse} uses the ledger-selected primary $C_1$, $C_2$, and $C_3$ runs and the ten $C_6$ runs, with run-bootstrap bands.

Within-group convergence grows over the episode in the two segregated interaction conditions. On Greenfield, separation moves from $+0.059$ to $+0.087$ in $C_2$ and from $+0.027$ to $+0.109$ in $C_3$ between rounds 5 and 30, and on Riverton from $+0.040$ to $+0.067$ and from $+0.019$ to $+0.061$. The solo-generation condition starts near the $C_3$ level because of its asymmetric briefs and then stays flat or declines ($+0.026$ to $+0.024$, and $+0.020$ to $+0.004$). The open-network $C_1$ stays at zero. Agents in segregated conditions talk only within their group, so part of this convergence is shared conversational vocabulary. The measure is descriptive, computed afterwards from stored probe text.

\section{Cross-variant tables and routing}\label{app:cross}\label{app:routing}

\begin{table}[h]
\centering\footnotesize
\setlength{\tabcolsep}{3.5pt}
\caption{\textbf{Cross-model $(C_2){-}(C_5)$:} $\bar\Delta$, $g$, and Welch $p$ per model and scenario. \yes: $p{<}0.05$. Same dir: sign matches the primary model.}\label{tab:cross}
\begin{tabular}{l ccc ccc c}
\toprule
& \multicolumn{3}{c}{Greenfield} & \multicolumn{3}{c}{Riverton} & Same dir \\
\cmidrule(lr){2-4}\cmidrule(lr){5-7}
Variant & $\bar{\Delta}$ & $g$ & $p$ & $\bar{\Delta}$ & $g$ & $p$ & GF, RV \\
\midrule
Qwen3-122B-FP8 (primary) & $+0.49$ & $+0.56$ & $0.007$\,\yes & $+0.65$ & $+0.62$ & $0.002$\,\yes & ref \\
Qwen3-VL-32B (text-only) & $+3.29$ & $+6.65^{\dagger}$ & $<10^{-7}$\,\yes & $+3.00$ & $+6.61^{\dagger}$ & $<10^{-7}$\,\yes & \yes, \yes \\
DeepSeek-V3 & $+0.08$ & $+0.10$ & $0.71$ & $+0.23$ & $+0.26$ & $0.36$ & \yes, \yes \\
GPT-4.1-mini & $-0.02$ & $-0.05$ & $0.87$ & $+0.38$ & $+0.83$ & $0.005$\,\yes & ${\approx}0$, \yes \\
Gemini-2.5-Flash & $+1.06$ & $+1.30$ & $<10^{-4}$\,\yes & $+1.50$ & $+2.08$ & $<10^{-7}$\,\yes & \yes, \yes \\
Llama-4-Scout (OpenRouter) & $+0.46$ & $+0.27$ & $0.34$ & $+0.35$ & $+0.27$ & $0.35$ & \yes, \yes \\
\bottomrule
\multicolumn{8}{l}{\footnotesize $^{\dagger}$Raw $\bar{\Delta}$ is the reported effect. $g$ is inflated because the small-$n$ pooled SD on its $C_5$ approaches the floor.} \\
\end{tabular}
\end{table}

\begin{table}[h]
\centering\footnotesize
\caption{\textbf{Cross-model $(C_2){-}(C_4)$:} run-level contrast per scenario and averaged, three runs per cell.}\label{tab:cross-c4}
\begin{tabular}{lrrr}
\toprule
Variant & Greenfield & Riverton & Scenario average \\
\midrule
DeepSeek-V3 & +0.292 & -0.042 & +0.125 \\
Gemini-2.5-Flash & +0.938 & +1.208 & +1.073 \\
GPT-4.1-mini & +0.125 & +0.062 & +0.094 \\
Llama-4-Scout & +0.104 & +0.542 & +0.323 \\
Qwen3-VL-32B & +3.167 & +3.083 & +3.125 \\
\bottomrule
\end{tabular}
\end{table}

The released analysis files give the cross-variant analogue of Table~\ref{tab:cross} for conformity and consistency, per scenario and variant. Conformity replication is mixed, with Gemini-2.5-Flash significant on Greenfield, GPT-4.1-mini significant on Riverton, and others mostly null. Consistency replication is broad, with 5 of 6 variants positive on Riverton, 3 of 6 on Greenfield, and no significant decrease. These $(C_2){-}(C_5)$ estimands are not pooled with the $(C_2){-}(C_4)$ panel.

Each additional-variant $C_2/C_4$ scenario cell contains three completed runs. Selection uses the last complete timestamped retry for each prescribed cell, as specified by the original cross-variant analysis, and all complete candidates remain listed. A completed cell with a reversed effect remains included. For DeepSeek-V3, historical $C_2$ configurations record \texttt{deepseek-chat}, while the selected $C_4$ runs record \texttt{deepseek-v4-flash}. Historical immutable revisions are not available, so we report the identifier difference without treating it as a verified same-revision experiment. Endpoint configuration used DeepSeek-V3, GPT-4.1-mini, Gemini-2.5-Flash, and Llama-4-Scout via provider APIs, with Llama-4-Scout routed through OpenRouter. Each model is used under its provider's API terms or its released model license. The primary analysis filters cross-variant replicas out of the within-Qwen3-122B-FP8 factorial via run-id pattern. The sign summary averages the two scenario differences within each variant before counting directions, so each variant counts once.

\subsection{Per-model interaction tests}\label{app:added}

To test each model on its own, we also run $(C_2){-}(C_4)$ on five API-served variants with 8 $C_2$ and 5 $C_4$ runs per scenario, as for the primary model (Table~\ref{tab:added}). The selection rule, seeds, and analysis follow a fixed protocol released with the code. For GPT-4.1-mini and Llama-4-Scout, these cells include the three runs per condition above, with the same configurations. DeepSeek-V4.1-Flash, Seed-2.0-mini, and GLM-4.7-Flash run with reasoning off. DeepSeek-V4.1-Flash is served under a rolling model name and is kept separate from DeepSeek-V3. The table shows each model's $C_4$ drift, its no-interaction retest level, next to the difference. GLM-4.7-Flash sits near the random-answer level: in a post hoc check, its retest L1 for an identical prompt is 4.85, against 5.67 for uniform random answers and 0.55 for GPT-4.1-mini, so its cells are uninformative about interaction.

\begin{table}[h]
\centering\footnotesize
\setlength{\tabcolsep}{4pt}
\caption{\textbf{Per-model $(C_2){-}(C_4)$ tests:} run-mean BFI-10 drift per condition, difference, 95\% run-bootstrap CI (50,000 draws), and two-sided exact permutation $p$.}\label{tab:added}
\begin{tabular}{llcccccc}
\toprule
Variant & Scenario & Runs $C_2/C_4$ & $C_4$ & $C_2$ & $\bar{\Delta}$ & 95\% CI & $p$ \\
\midrule
GPT-4.1-mini & Greenfield & 8/5 & 0.39 & 0.50 & $+0.112$ & $[+0.01, +0.22]$ & 0.11 \\
 & Riverton & 8/5 & 0.70 & 0.70 & $+0.003$ & $[-0.14, +0.14]$ & 1.00 \\
\addlinespace[2pt]
Llama-4-Scout & Greenfield & 8/5 & 1.23 & 1.76 & $+0.533$ & $[+0.01, +0.99]$ & 0.11 \\
 & Riverton & 8/5 & 0.50 & 1.62 & $+1.125$ & $[+0.74, +1.54]$ & 0.0008 \\
\addlinespace[2pt]
DeepSeek-V4.1-Flash & Greenfield & 8/5 & 1.55 & 3.20 & $+1.645$ & $[+1.27, +2.02]$ & 0.0008 \\
 & Riverton & 8/5 & 1.69 & 2.98 & $+1.289$ & $[+0.96, +1.61]$ & 0.0008 \\
\addlinespace[2pt]
Seed-2.0-mini & Greenfield & 8/5 & 1.59 & 2.22 & $+0.631$ & $[+0.44, +0.81]$ & 0.0008 \\
 & Riverton & 8/5 & 1.00 & 1.84 & $+0.836$ & $[+0.63, +1.07]$ & 0.0008 \\
\addlinespace[2pt]
GLM-4.7-Flash & Greenfield & 8/5 & 4.15 & 4.26 & $+0.108$ & $[-0.48, +0.75]$ & 0.81 \\
 & Riverton & 8/5 & 4.15 & 4.45 & $+0.295$ & $[-0.29, +0.92]$ & 0.48 \\
\bottomrule
\end{tabular}
\end{table}

\section{Signed and temperature robustness}\label{app:signed}

\begin{table}[h]
\centering\footnotesize
\caption{\textbf{Signed factors:} $\kappa_f$, 95\% run-bootstrap CI (10,000 draws), and replica agreement per factor. Agreement: $C_2$ runs on the side of $\kappa_f$, out of 8.}\label{tab:signed}
\begin{tabular}{llrrc}
\toprule
Scenario & Factor & $\kappa_f$ & 95\% interval & Agreement \\
\midrule
Greenfield & Extraversion & +0.302 & $[+0.095,+0.481]$ & 8/8 \\
Greenfield & Agreeableness & +0.014 & $[-0.098,+0.125]$ & 3/8 \\
Greenfield & Conscientiousness & +0.072 & $[+0.003,+0.142]$ & 7/8 \\
Greenfield & Neuroticism & +0.358 & $[+0.227,+0.505]$ & 8/8 \\
Greenfield & Openness & +0.072 & $[-0.145,+0.277]$ & 6/8 \\
Riverton & Extraversion & -0.084 & $[-0.269,+0.106]$ & 5/8 \\
Riverton & Agreeableness & -0.263 & $[-0.400,-0.145]$ & 8/8 \\
Riverton & Conscientiousness & +0.016 & $[-0.025,+0.059]$ & 1/8 \\
Riverton & Neuroticism & +0.053 & $[-0.091,+0.202]$ & 5/8 \\
Riverton & Openness & -0.362 & $[-0.514,-0.195]$ & 7/8 \\
\bottomrule
\end{tabular}
\end{table}

\begin{figure}[h]
\centering
\includegraphics[width=0.92\textwidth]{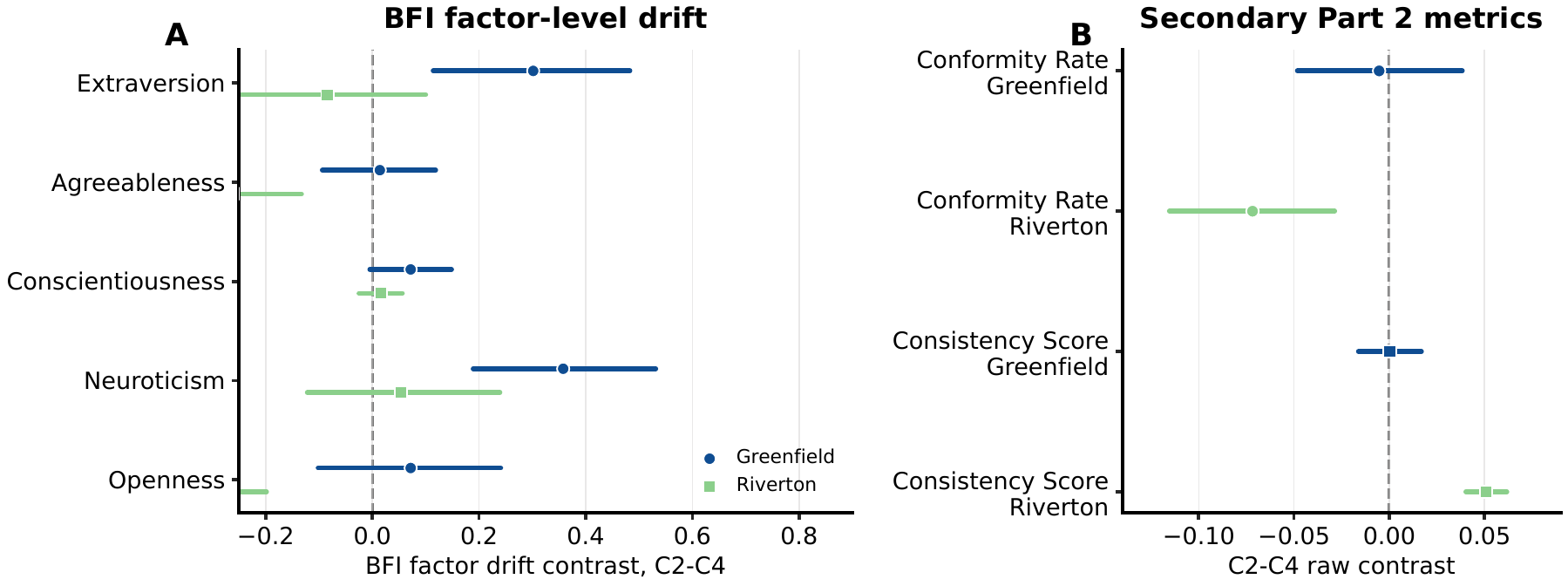}
\caption{\textbf{Part 2 breakdown:} BFI-10 factor-level drift, conformity and consistency.}\label{fig:app-part2-breakdown}
\end{figure}

The deterministic re-probe uses the same ten BFI items and reverse-coding rules, replacing only probe temperature. Saved post-treatment states retain their history, so it is a sensitivity analysis over the same experimental trajectories rather than an independently generated replication. Factor analyses are descriptive and uncorrected for multiple comparisons such as false-discovery-rate control~\citep{Benjamini1995}. Figure~\ref{fig:app-part2-breakdown} gives the original factor-level drift and secondary Part 2 metrics.

\section{Supporting analyses}\label{app:secondary}

\begin{table}[h]
\centering\footnotesize
\caption{\textbf{No-SRM check:} $C_2$ and $C_5$ means, $\bar\Delta$ ($g$), CI and Welch $p$; 3 vs.\ 3 runs.}\label{tab:nosrm}
\resizebox{\textwidth}{!}{%
\begin{tabular}{lcccc}
\toprule
Scenario & $C_2$ mean & $C_5$ mean & $\bar{\Delta}$ ($g$) & 95\% CI, $p$ \\
\midrule
Greenfield & $2.48$ & $0.77$ & $+1.71$ ($+2.02$) & $[+1.04,+2.48]$, $p{=}5.9\times10^{-8}$ \\
Riverton & $1.85$ & $0.90$ & $+0.96$ ($+0.94$) & $[+0.23,+1.81]$, $p{=}0.0019$ \\
\bottomrule
\end{tabular}}
\end{table}

No-SRM $C_2$ run means are $[2.000,2.250,3.188]$ in Greenfield and $[1.500,2.500,1.563]$ in Riverton, against $C_5$ means $[0.938,0.688,0.688]$ and $[1.125,1.063,0.500]$. Exact two-sided tests enumerate $\binom{6}{3}=20$ allocations and give $p=0.10$ in both cases, so the agent-level Welch values in Table~\ref{tab:nosrm} serve as descriptive evidence, and the run-level exact test carries the inference.

\begin{table}[h]
\centering\footnotesize
\begin{minipage}[t]{0.47\textwidth}\centering
\caption{\textbf{Secondary instruments:} $C_2-C_4$ difference, 95\% CI.}\label{tab:secondary}
\resizebox{\linewidth}{!}{%
\begin{tabular}{llrr}
\toprule
Scen. & Instrument & Diff. & 95\% CI \\
\midrule
Greenfield & Conformity & -0.005 & $[-0.048,+0.038]$ \\
Greenfield & Wording consistency & +0.000 & $[-0.016,+0.017]$ \\
Riverton & Conformity & -0.072 & $[-0.115,-0.029]$ \\
Riverton & Wording consistency & +0.051 & $[+0.040,+0.062]$ \\
\bottomrule
\end{tabular}}
\end{minipage}\hfill
\begin{minipage}[t]{0.5\textwidth}\centering
\caption{\textbf{Allocation by dimension:} $C_2-C_4$ in pp, 95\% CI.}\label{tab:allocation}
\resizebox{\linewidth}{!}{%
\begin{tabular}{llrr}
\toprule
Scen. & Dimension & Diff. & 95\% CI \\
\midrule
Greenfield & Education & +7.09 & $[+1.53,+13.39]$ \\
Greenfield & Healthcare & +4.82 & $[+2.47,+8.13]$ \\
Greenfield & Green Spaces & +0.78 & $[-1.06,+2.84]$ \\
Greenfield & Infrastructure & +6.10 & $[+3.44,+9.94]$ \\
Greenfield & Economic Development & +3.90 & $[+2.14,+7.38]$ \\
Riverton & Automation & +4.30 & $[+1.97,+9.24]$ \\
Riverton & Workforce Training & +4.09 & $[+2.78,+6.61]$ \\
Riverton & Employee Support & -1.70 & $[-2.72,+0.66]$ \\
Riverton & Innovation Lab & +5.69 & $[+3.09,+9.19]$ \\
Riverton & Data Infrastructure & +0.22 & $[-1.97,+3.30]$ \\
\bottomrule
\end{tabular}}
\end{minipage}
\end{table}

\clearpage
\section*{NeurIPS Paper Checklist}

\begin{enumerate}

\item {\bf Claims}
    \item[] Question: Do the main claims made in the abstract and introduction accurately reflect the paper's contributions and scope?
    \item[] Answer: \answerYes{}
    \item[] Justification: Sections~\ref{sec:intro}--\ref{sec:exp-limits} state the tested scope, raw effects, run-level inference, and the matched solo-generation boundary.

\item {\bf Limitations}
    \item[] Question: Does the paper discuss the limitations of the work performed by the authors?
    \item[] Answer: \answerYes{}
    \item[] Justification: Section~\ref{sec:exp-limits} discusses construct, temporal, matching, sample-size, model-revision, and generalization limits.

\item {\bf Theory assumptions and proofs}
    \item[] Question: For each theoretical result, does the paper provide the full set of assumptions and a complete (and correct) proof?
    \item[] Answer: \answerNA{}
    \item[] Justification: The work reports empirical experiments and descriptive statistical estimands, not theoretical results.

    \item {\bf Experimental result reproducibility}
    \item[] Question: Does the paper fully disclose all the information needed to reproduce the main experimental results of the paper to the extent that it affects the main claims and/or conclusions of the paper (regardless of whether the code and data are provided or not)?
    \item[] Answer: \answerYes{}
    \item[] Justification: Sections~\ref{sec:method} and~\ref{sec:fsip-cycle} and Appendix~\ref{app:materials} give the protocol, prompts, configurations, seeds, and decoding settings, and the released measurements and scripts reproduce every reported analysis offline. Serving metadata are described in Appendix~\ref{app:audit}.

\item {\bf Open access to data and code}
    \item[] Question: Does the paper provide open access to the data and code, with sufficient instructions to faithfully reproduce the main experimental results, as described in supplemental material?
    \item[] Answer: \answerYes{}
    \item[] Justification: Code, selected measurements, hashes, and offline reproduction instructions are public at \url{https://github.com/koanatakiyo/SocCogEval}, with code under the MIT license and data under CC BY 4.0.

\item {\bf Experimental setting/details}
    \item[] Question: Does the paper specify all the training and test details (e.g., data splits, hyperparameters, how they were chosen, type of optimizer) necessary to understand the results?
    \item[] Answer: \answerYes{}
    \item[] Justification: Sections~\ref{sec:method} and~\ref{sec:fsip-cycle} specify conditions, agent counts, rounds, probes, temperatures, and contrasts. The frozen C6 protocol and serving amendment are included.

\item {\bf Experiment statistical significance}
    \item[] Question: Does the paper report error bars suitably and correctly defined or other appropriate information about the statistical significance of the experiments?
    \item[] Answer: \answerYes{}
    \item[] Justification: Section~\ref{sec:metrics} defines run-level exact tests and descriptive bootstrap intervals. Secondary estimates and small-cell limitations are labeled separately.

\item {\bf Experiments compute resources}
    \item[] Question: For each experiment, does the paper provide sufficient information on the computer resources (type of compute workers, memory, time of execution) needed to reproduce the experiments?
    \item[] Answer: \answerYes{}
    \item[] Justification: Appendix~\ref{app:cost} reports the GPU type and GPU-hours for the locally served runs, and the released artifact tables give call and token totals for the API-served and $C_6$ runs.
    
\item {\bf Code of ethics}
    \item[] Question: Does the research conducted in the paper conform, in every respect, with the NeurIPS Code of Ethics \url{https://neurips.cc/public/EthicsGuidelines}?
    \item[] Answer: \answerYes{}
    \item[] Justification: The study uses synthetic scenarios and LLM agents, without human participants or personal participant data. Construct limits and intended use are discussed in the paper.

\item {\bf Broader impacts}
    \item[] Question: Does the paper discuss both potential positive societal impacts and negative societal impacts of the work performed?
    \item[] Answer: \answerYes{}
    \item[] Justification: Section~\ref{sec:exp-limits} describes the uncertainty-estimation benefit and cautions against treating simulated profiles as measurements of real people.
    
\item {\bf Safeguards}
    \item[] Question: Does the paper describe safeguards that have been put in place for responsible release of data or models that have a high risk for misuse (e.g., pre-trained language models, image generators, or scraped datasets)?
    \item[] Answer: \answerNA{}
    \item[] Justification: The release consists of experiment code and synthetic-agent measurements, not pretrained weights, scraped personal data, or a high-risk generative model.

\item {\bf Licenses for existing assets}
    \item[] Question: Are the creators or original owners of assets (e.g., code, data, models), used in the paper, properly credited and are the license and terms of use explicitly mentioned and properly respected?
    \item[] Answer: \answerYes{}
    \item[] Justification: Models and instruments are cited to their sources. Each model is used under its provider's API terms or its released model license, and the instruments are used for research only (Appendix~\ref{app:routing}). The release documents third-party dependencies and attributions. Our code is released under MIT and our data under CC BY 4.0.

\item {\bf New assets}
    \item[] Question: Are new assets introduced in the paper well documented and is the documentation provided alongside the assets?
    \item[] Answer: \answerYes{}
    \item[] Justification: Appendix~\ref{app:materials} and the accompanying README and data documentation describe the code, measurements, manifests, analysis commands, and reproducibility boundaries.

\item {\bf Crowdsourcing and research with human subjects}
    \item[] Question: For crowdsourcing experiments and research with human subjects, does the paper include the full text of instructions given to participants and screenshots, if applicable, as well as details about compensation (if any)? 
    \item[] Answer: \answerNA{}
    \item[] Justification: There were no human participants or crowdsourcing workers.

\item {\bf Institutional review board (IRB) approvals or equivalent for research with human subjects}
    \item[] Question: Does the paper describe potential risks incurred by study participants, whether such risks were disclosed to the subjects, and whether Institutional Review Board (IRB) approvals (or an equivalent approval/review based on the requirements of your country or institution) were obtained?
    \item[] Answer: \answerNA{}
    \item[] Justification: The experiments involve synthetic LLM agents and no human subjects.

\item {\bf Declaration of LLM usage}
    \item[] Question: Does the paper describe the usage of LLMs if it is an important, original, or non-standard component of the core methods in this research? Note that if the LLM is used only for writing, editing, or formatting purposes and does \emph{not} impact the core methodology, scientific rigor, or originality of the research, declaration is not required.
    \item[] Answer: \answerYes{}
    \item[] Justification: LLMs are the experimental agents. Their roles, recorded model identifiers, prompts, decoding settings, and routing limitations are described in Sections~\ref{sec:method} and~\ref{sec:fsip-cycle} and Appendix~\ref{app:routing}.

\end{enumerate}
\end{document}